\documentclass[aps,prx,onecolumn,amsmath,nofootinbib,amssymb,11pt]{revtex4-1}
\usepackage[T1]{fontenc}
\usepackage[latin9]{luainputenc}
\usepackage[T1]{fontenc}
\usepackage[latin9]{luainputenc}
\usepackage{amsmath,bm}
\usepackage{fancyhdr}
\usepackage{amssymb}
\usepackage[colorlinks, linkcolor=blue]{hyperref}
\usepackage{graphicx}
\usepackage[format=hang,justification=centerlast]{subcaption}
\usepackage[normalem]{ulem}
\usepackage[export]{adjustbox}
\usepackage{soul}      
\usepackage{color}
\renewcommand{\vec}[1]{\boldsymbol{#1}}

\def \ve {\varepsilon}
\def \r {{\vec r}}

\def \v {{\bf v}}

\def \kb {k_\tn{B}}

\def \ve {\varepsilon}

\def \beq {\begin{eqnarray}}
\def \eeq {\end{eqnarray}}
\def \tn {\textnormal}

\def \la{\langle}
\def \ra{\rangle}

\def \hb {H_{\tn{B}}}
\def \hc {H_{\tn{c}}}

\def \rd {\rho_{\tn{dc}}}
\def \Ht {H_{\textnormal{kin,c}}}
\def \Hc {H_{\textnormal{int,c}}}

\def \Htb {H_{\textnormal{kin,B}}}
\def \Hcb {H_{\textnormal{int,B}}}

\makeatletter
\usepackage{braket}
\usepackage{graphicx}

\makeatother

\usepackage[english]{babel}

\begin{document}
\title{\textsf{Bad metallic transport in geometrically frustrated models}}
\sffamily
\author{\textsf{J. F. Mendez-Valderrama}}
\author{\textsf{Debanjan Chowdhury}}

\affiliation{\textsf{Department of Physics, Cornell University, Ithaca, New York 14853, USA.}}

\begin{abstract}
We study the transport properties for a family of geometrically frustrated models on the triangular lattice with an interaction scale far exceeding the single-particle bandwidth. Starting from the interaction-only limit, which can be solved exactly, we analyze the transport and thermodynamic behavior as a function of filling and temperature at the leading non-trivial order in the single-particle hopping. Over a broad range of intermediate temperatures, we find evidence of a dc resistivity scaling linearly with temperature and with typical values far exceeding the quantum of resistance, $h/e^2$. At a sequence of commensurate fillings, the bad-metallic regime eventually crosses over into interaction induced insulating phases in the limit of low temperatures. We discuss the relevance of our results to experiments in cold-atom and  moir\'e heterostructure based platforms.  
\end{abstract}

\maketitle

\noindent\rule{\textwidth}{1pt}
\tableofcontents
\noindent\rule{\textwidth}{1pt}

\section{\textsf{Introduction}}
\label{sec:intro}

Electrical transport in weakly interacting metals is well described by Boltzmann theory \cite{ziman}, where the finite resistance arises due to scattering of well defined, long-lived electronic (``Landau'') quasiparticles. At low-temperatures, the relevant quasiparticle excitations that participate in transport in a Fermi liquid are restricted to the near vicinity of a sharp Fermi surface. There are numerous correlated materials with metallic states that exhibit a variety of non-Fermi liquid (NFL) properties. Well known examples of such behavior have been documented in the iron-based superconductors \cite{scalapino,Matsuda14}, rare-earth element based materials \cite{Stewart,rmpqcp}, ruthenates \cite{allen,Tyler1998,gegenwart}, cobaltates \cite{Taillefer1,Ong1}, and perhaps most notably in the copper-oxide based compounds \cite{Takagi,Keimer15}. A theoretical description of how the single-particle and transport lifetimes evolve as one departs from the weak-coupling regime remains an area of active research.  

In spite of the rich and diverse phenomenology across the different NFL metals, some of the striking universal features include: (i) unusual power-laws in dc transport  \cite{Takagi,allen,Tyler1998,gegenwart,Taillefer1,Ong1} and optical conductivity  \cite{CROopt,Marel03} as a function of temperature and frequencies, respectively, that are inconsistent with the expectations in a Fermi liquid, and, (ii) values of resistivity far exceeding the Mott-Ioffe-Regel (MIR) limit, where the electronic mean free path becomes comparable to the lattice spacing, without any sign of a saturation upto the highest measurable temperatures  \cite{Gunnarsson,Hussey04}. The smooth evolution of ``bad-metallic'' behavior \cite{SAK95} down to low temperatures without any characteristic crossovers necessitates a single theoretical framework that can capture the transport behavior over the entire range of temperatures. A recent cold-atoms based experiment studying the doped Fermi-Hubbard model on the square lattice has also observed bad-metallic behavior with a $T-$linear resistivity \cite{Bakr} over a reasonably broad range of temperatures, when the interaction scale is comparable to the electronic bandwidth.

Empirically, it is clear that there are different routes towards realizing NFL physics in the panoply of materials introduced above. Certain families of correlated materials, e.g. the ruthenates \cite{allen,Tyler1998,gegenwart} and cobaltates \cite{Taillefer1,Ong1}, display bad-metallic and NFL behavior over a wide, intermediate range of temperatures. This regime is often accompanied by a resistivity that scales linearly with the temperature, $\rho_{\tn{dc}}\sim T$, and a universal ``Drude'' scattering rate that is independent of any microscopic detail \cite{Bruin13}. Below a certain characteristic coherence temperature, $T_{\tn{coh}}$, there is a crossover to conventional FL behavior where electronic quasiparticles emerge as well-defined low-energy excitations. Interestingly, the zero-temperature entropy extrapolated from the intermediate scale NFL regime ($T>T_{\tn{coh}}$) for some of these materials is finite \cite{allen,bruhwiler}, and is relieved below $T_{\tn{coh}}$. In this paper, we will similarly focus on a class of ``infrared (IR) incomplete'' metallic states \cite{DC2018}, where the bad-metallic behavior arises over a broad range of temperatures, $T_{\tn{IR}} < T < T_{\tn{UV}}$. One of the defining features of the IR-incomplete states is that $T_{\tn{IR}}$ can not be tuned to be zero, i.e. the intermediate scale properties are not controlled by a $T=0$ IR renormalization-group fixed point with a few relevant perturbations. 

It is natural to ask if there are ``solvable'' models, which can be analyzed reliably in the regime of strong interactions, that lead to IR-incomplete bad-metal behavior accompanied by $T-$linear resistivity. In this paper, we explore the transport properties of a family of geometrically frustrated models on the triangular lattice. In the strong-coupling limit, the problem effectively reduces to the Ising model on the triangular lattice which can be analyzed using classical Monte-Carlo simulations and the transport coefficients can be computed perturbatively in the hopping strength \cite{Hartnoll18}. Over an intermediate range of temperatures, these models display bad-metallic behavior that can not be described by a conventional Landau-Boltzmann quasiparticle-based paradigm. Moreover, significant analytical insight into the nature of this regime can be gained from the exact solution for the Ising model on the triangular lattice due to Wannier \cite{WannierTriaglat,WannierTriaglat2}.

An alternative approach, that has already shed a lot of interesting light on a class of IR-incomplete NFLs and has been successful in capturing some of the NFL phenomenology over a broad range of temperature and energy scales, utilizes the Sachdev-Ye-Kitaev (SYK) model \cite{SY,kitaev_talk,Parcollet1,Parcollet2,Maldacena2016} as a solvable building block. The SYK model is a $(0+1)-$dimensional model consisting of $M$ orbitals interacting with random all-to-all (frustrated) interactions on a single site. A number of subsequent works have constructed higher-dimensional lattice generalizations of the SYK model and studied their transport properties \cite{Gu17,SS17,shenoy,Balents,Zhang17,mcgreevy,DVK17,Yao,SSmagneto,DC2018}. A particular class of models studied in these constructions involve a single-band with a bandwidth, $W$, and typical interaction strength $U$. In the limit of $M$ large, the system exhibits a crossover from a low-temperature Landau Fermi-liquid to an incoherent NFL at high temperatures without a remnant Fermi surface \cite{Parcollet1,Balents,DC2018}. Interestingly, the NFL state at $T>T_{\tn{coh}}(\sim W^2/U)$ exhibits a DC resistivity, $\rho_{\tn{dc}}=(h/Me^2) T/T_{\tn{coh}}$, accompanied by a finite residual entropy if one extrapolates to $T\rightarrow0$, in addition to other appealing features \cite{DC2018}. The geometrically frustrated models we study in this paper share some resemblance with the results obtained for the SYK models, but notably have significant differences in the origin of the NFL physics, as we discuss later. 

The rest of this paper is organized as follows: In Sec.~\ref{sec:prelim}, we introduce a family of models on the triangular lattice, written in terms of either hard-core bosons, spin-1/2 moments or spinless electrons interacting via nearest-neighbor interactions, that we study in the strong interaction limit. In Sec.~\ref{subsec:pert} we describe our perturbative approach that involves deforming the original model by weak longer-ranged (``screened coulomb'') interactions, which leads to a number of interesting consequences. In the interaction-only limit, the deformed models are completely classical and in Sec.~\ref{sec:thermo}, we use classical Monte-Carlo simulations to evaluate the density-density correlation functions and the various thermodynamic coefficients, in addition to mapping out the full phase-diagram as a function of filling and temperature. In Sec.~\ref{sec:transport}, we investigate the transport properties of the above models to leading non-trivial order in the single electron hopping strength across the entire phase-diagram, extending all the way from the high-temperature thermally disordered regime to the low-temperature insulating regimes (at commensurate fillings) through an intermediate correlated metallic regime. We end with an outlook for interesting open questions in sec.~\ref{sec:outlook}, which includes a discussion of the possible interesting connections to future experiments in cold-atoms and moir\'e heterostructures based setups and sign-problem free quantum Monte-Carlo computations for one of the models introduced in the paper. We leave a discussion of some of the technical details to appendices \ref{ap:MC}-\ref{ap:Is}.

\section{\textsf{A family of models}}
\label{sec:prelim}

\subsection{\textsf{Hard-core bosons on triangular lattice}}
\label{subsec:HB}
Consider the following Hamiltonian for hard-core, charge $e$ bosons (with a globally conserved density), defined on the sites ($\r$) of a triangular lattice,
\begin{subequations}
\beq
\hb &=& \Htb + \Hcb \label{Hb}\\
\Htb &=& -t\sum_{\la\r,\r'\ra} ~b_\r^\dagger b_{\r'} + \tn{H.c.} - \mu_b \sum_\r n_\r^b, \label{Htb}\\
\Hcb &=& V\sum_{\la\r,\r'\ra}~ \left(n^b_\r - \frac{1}{2}\right) \left(n^b_{\r'} - \frac{1}{2}\right),
\eeq
\end{subequations}
where $b_\r^\dagger$ and $b_\r$ denote the bosonic creation and annihilation operators. The nearest-neighbor hopping and interaction strengths are denoted $t$ and $V$, respectively. The hard-core boson density at a site $\r$ is given by $n^b_\r = b_\r^\dagger b_\r~(=0,~1)$. The average density can be tuned by varying the chemical potential, $\mu_b$.

Before we describe some of the properties of $\hb$ in Eqn.~\ref{Hb}, it is worth noting that at half-filling the above model also maps to the standard spin-1/2 XXZ model: 
\beq
H_{\tn{XXZ}} &=& \sum_{\la\r,\r'\ra} \bigg[-J_\perp(S_\r^x S_{\r'}^x + S_\r^y S_{\r'}^y) + J_z S_\r^z S_{\r'}^z \bigg], 
\label{Hxxz}
\eeq
where $J_z \equiv V$, $J_\perp \equiv 2t$ and $S_\r^\alpha$ ($\alpha=x,y,z$) represent the standard spin-1/2 operators at site, $\r$. The pure antiferromagnetic (AFM) Ising limit in the above model corresponds to $J_\perp=0$ and  $J_z>0$; in Eqn.~\ref{Hb} this limit corresponds to bosons without any kinetic energy and repulsive  nearest-neighbor interactions.  The thermodynamic properties of the classical Ising limit on the triangular lattice are well understood, in part due to the exact solutions \cite{WannierTriaglat,WannierTriaglat2}. In the absence of external fields (i.e. at half-filling of the bosonic model), the system lacks ordering down to $T=0$ with algebraic spin correlations \cite{StephensonCorr} and a residual entropy of $S/N \approx 0.323 ~\kb$ \cite{HOUTAPPEL1950}, with $N$ being the number  of sites.

Quantum fluctuations can be included in the above models by perturbing around the classical (Ising) limit by introducing boson hopping terms (Eqn.~\ref{Htb}). The resulting ground state phase diagram for $\hb$ has been studied extensively at low temperatures using quantum Monte-Carlo calculations \cite{melko,troyer,damle}, as a function of filling, $n$, and $\lambda=V/t$. The model exhibits superfluidity, with a finite superfluid stiffness, up to large values of $\lambda$. Additionally, for a range of densities $1/3\leq n\leq 2/3$ and $\lambda\gtrsim 1$, the ground state exhibits translational symmetry breaking leading to supersolid order via an `order-by-disorder' mechanism. 

While the equilibrium correlation functions, such as the superfluid stiffness and the static structure factor, have been studied in great detail for the above model, the dynamical properties at a finite temperature remain poorly understood. In particular, the finite temperature transport properties, especially at temperatures above the scale associated with the superfluid stiffness, are not known. Since the bosons become coherent only below this scale, it is possible that in the strong coupling limit ($\lambda\gg1$) and over a broad range of temperatures above the superfluid stiffness, the above model realizes an incoherent metal with unconventional transport properties. 

Inspired by these interesting observations, we shall study the possibility of incoherent transport in a different (but conceptually related) electronic model in the next subsection \ref{subsec:sf}, where we address the transport properties in a solvable setting. Our calculations will be performed in a regime where the hopping parameter, $t$, is treated perturbatively. We expect the statistics of the underlying excitations to not play an essential role in this regime and therefore, the transport properties of the bosonic and electronic models should be essentially identical. We note that an example of a bad metallic regime was pointed out in an unrelated hard-core boson model in Ref.~\cite{lindner}.

\subsection{\textsf{Spinless fermions on triangular lattice}}
\label{subsec:sf}

Let us now introduce a model of spinless fermions (or fully spin-polarized electrons) on the triangular lattice,
\begin{subequations}
\beq
\hc &=& \Ht + \Hc \label{Hc}\\
\Ht &=& -t\sum_{\la\r,\r'\ra }\left(c_{\r}^{\dagger}c_{\r'} + \tn{H.c.}\right) - \mu_c \sum_\r n_\r^c, \label{Htc}\\
\Hc &=& V\sum_{\la\r,\r'\ra} \left(n^c_\r - \frac{1}{2}\right) \left(n^c_{\r'} - \frac{1}{2}\right),
    \label{eqn:hamiltonian}
\eeq
\end{subequations}
where the parameters of the model are as introduced earlier. The fermion density operator at $\r$ is denoted $n_{\r}^c=c^{\dagger}_{\r}c_{\r}~(=0,~1)$ and $\mu_c$ is the chemical potential that tunes the global density. The operators $ c_{\r},c_{\r}^\dagger$ denote the annihilation and creation operators, respectively. We note that the dimension of the local Hilbert space for all of the models introduced thus far, and, the interaction-only (classical) limit are identical. On the other hand, the nature of the ground-state and the asymptotically low-temperature properties are qualitatively distinct for the two classes of bosonic vs. fermionic models. 

At weak-coupling, $\lambda(=V/t)\ll 1$, the ground state of $\hc$ in Eqn.~\ref{Hc} is a Fermi liquid with a sharply defined Fermi surface. In this limit, the interaction strength leads to mere quantitative renormalizations of various thermodynamic coefficients. Since the quasiparticles near the Fermi surface are long-lived, DC transport can be analyzed using a conventional Boltzmann equation approach; in the absence of disorder, the only mechanism relevant for degrading momentum is via Umklapp scattering. For a range of intermediate densities (i.e. when the Fermi surface size occupies a significant portion of the Brillouin zone), the dc resistivity has a classic $T^2$ behavior and the parametric dependence can be expressed simply as, $\rho_{\tn{dc}}\sim (h/e^2) \overline{V}^2 (T/T_F)^2$, where $\overline{V}$ is a dimensionless interaction strength and $T_F$ is the Fermi-temperature. The values of the dc resistivity are thus much smaller than the quantum of resistance, $h/e^2$, in the regime where the above prescription is valid.

On the other hand, the ground-state properties of $\hc$ remain partly unknown at strong-coupling, $\lambda\gg1$. The model suffers from the infamous ``sign-problem'', unlike the bosonic version of the model ($\hb$) in Eqn.~\ref{Hb}, thereby preventing an exact numerical analysis using quantum Monte-Carlo simulations. Nevertheless, based on the ground-state properties of $\hb$, we can speculate that $\hc$ likely hosts (i) an interacting Fermi liquid ground-state with a finite quasiparticle residue upto large values of $\lambda$, and, (ii) additional density-wave correlations in the presence of metallicity for at least a range of commensurate fillings near $1/3<n<2/3$. {\footnote{\textsf{One may choose to use a ``parton'' decomposition: $c_\r = b_\r f_\r$, where $b_\r$ is an electrically charged hard-core boson and $f_\r$ is an electrically neutral fermion that forms a Fermi surface. The superfluid state of $b_\r$ then corresponds to a FL of $c_\r$ and the additional translational symmetry breaking associated with the $b-$supersolid \cite{melko,troyer,damle} can lead to a metallic density-wave. We do not elaborate on this line of thinking any further.}}} Earlier work on the electronic triangular lattice models have indeed used variational approaches with Gutzwiller projected, Jastrow-type wavefunctions and other numerical techniques to obtain a similar ground-state phase diagram \cite{MotrunichLee2003VMC, Tocchio,cryst2031155}.


We are interested in the fate of the metallic FL and its transport properties at strong-coupling, as a function of increasing temperatures. Just as the superfluid stiffness defined a scale above which the bosons are no longer coherent in the model defined by $\hb$ in Eqn.~\ref{Hb}, we anticipate that the fermions have a characteristic coherence scale above which a quasiparticle-based framework will no longer be sufficient to describe the transport properties. The remainder of this paper will focus on the thermodynamic and transport properties for the model defined in Eqn.~\ref{Hc} (with a minor technical modification to be described in Sec.~\ref{subsec:pert} below) over a broad intermediate range of temperatures and fillings.

\section{\textsf{Perturbative approach from Ising limit}}
\label{subsec:pert}

\begin{figure*}[t!]
    \begin{subfigure}[t]{0.50\textwidth}
        \includegraphics[width=0.85\linewidth]{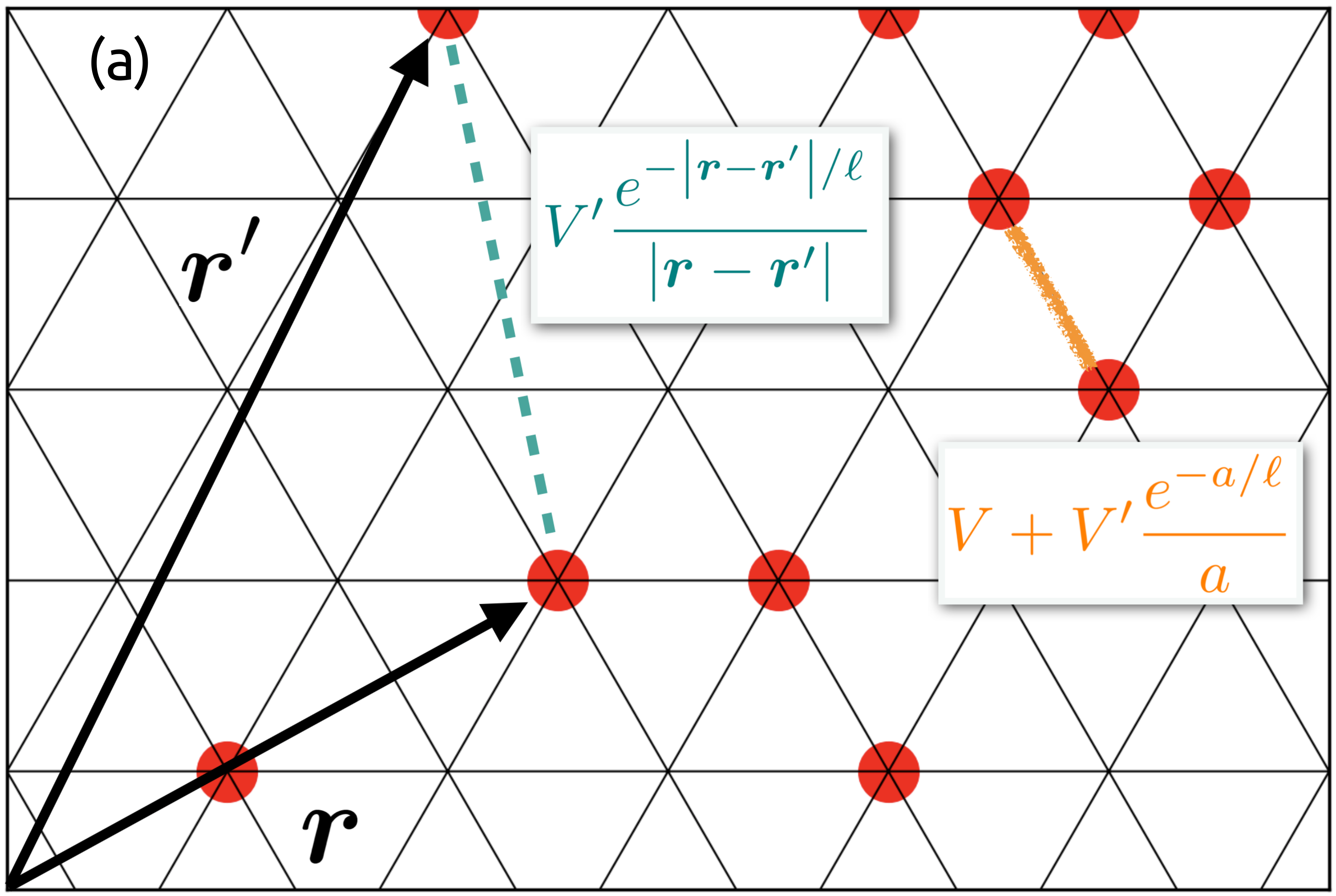}
    \end{subfigure}%
    ~ 
    \begin{subfigure}[t]{0.50\textwidth}
        \includegraphics[width=0.85\linewidth]{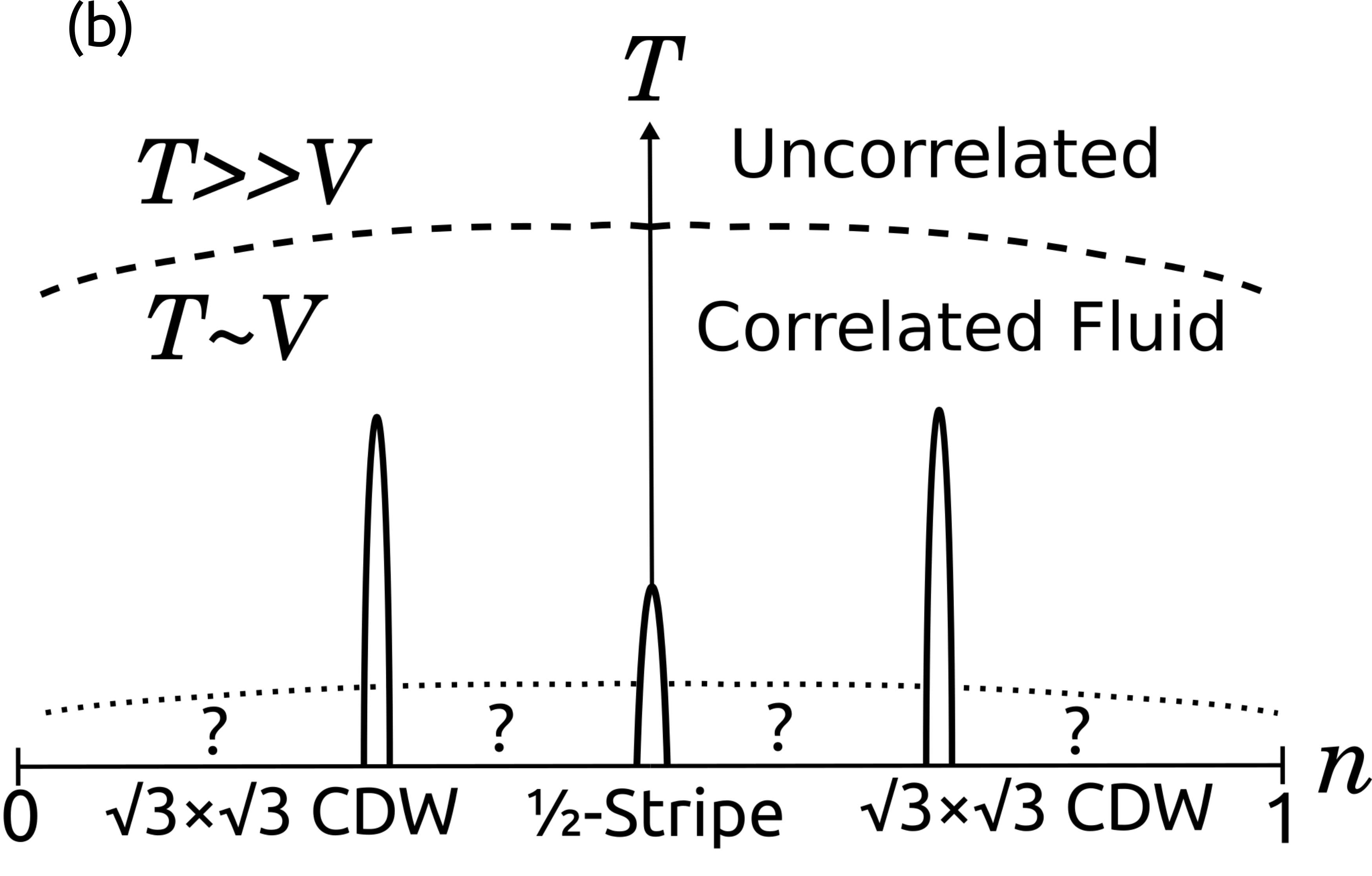}
    \end{subfigure}
    \caption{
        \textsf{(a) An illustration of the relevant interaction terms in the modified Hamiltonian, $\Hc+\delta\Hc$, for a particular configuration. Note that $\delta\Hc$ modifies the nearest-neighbour interaction term (orange line). (b) A schematic phase-diagram as a function of temperature ($T$) and filling ($0<n<1$). The nearest-neighbour (Ising) interaction drives a transition to a $\sqrt{3}\times\sqrt{3}$ charge density wave (or generalized Wigner crystal) at fillings, $n=1/3,~2/3$ below $T\sim O(V)$. For $V'\neq0$, additional correlated insulating states can appear at a sequence of commensurate fractions, $n=p/q$, at lower temperatures.}}
    \label{fig:fig1} 
\end{figure*}

In order to analyze the transport properties of the model defined by $\hc$ in Eqn.~\ref{Hc} for $\lambda\gg1$, we can set up a perturbative treatment around the $t=0$ limit in powers of $1/\lambda$. In the $t=0$ (antiferromagnetic Ising) limit, $V$ plays the role of an exchange coupling and $\mu_c$ is equivalent to a Zeeman coupling to an external field. This limit has a massive degeneracy and the perturbative treatment for a finite (small) $t$ can only proceed via degenerate perturbation theory. Instead of taking this route, which we leave for future work, here we proceed by adopting an alternative procedure \cite{Hartnoll18}. We deform the original Hamiltonian by a small amount and as a result, lift the massive degeneracy. This allows us to proceed by adopting standard perturbation theory, without affecting the transport properties of the underlying model in a dramatic fashion. The additional deformation is inspired by a screened Coulomb interaction of the form 
\beq
\hc &\rightarrow& \hc + \delta\Hc,\\
\delta\Hc &=& \frac{V'}{2} \sum_{\r\neq\r'}\frac{\exp{\left( -|\r-\r'|/\ell\right) }}{|\r-\r'|}n^c_\r n^c_{\r'},
    \label{eqn:LongV}
\eeq
where $0<V'\ll V$ sets the overall scale of the perturbation and $\ell$ is the screening length; see Fig.~\ref{fig:fig1} (a). While the particular choice of the deformed density-density interaction likely does not affect the results significantly, the modification introduced above is physically motivated by the screened interaction present in a number of interesting moir\'e setups, including bilayers of transition metal dichalcogenides (TMD) in close proximity to a metallic gate. These longer ranged interactions play a crucial role in the TMD moir\'e setup on the triangular lattice in the limit of low temperatures \cite{xu2020abundance}, as we shall return to later. For the sake of convenience, we choose $\ell=2a$ for the remainder of our discussion, with $a$ the lattice constant. In particular, for $T\gg t$, we approximate $e^{-\beta H_c}\approx e^{-\beta(\Hc + \delta\Hc)}$, where $\beta=1/k_BT$, and the thermal ensemble for the resulting model can be analyzed reliably using classical Monte-Carlo calculations. In what follows, we drop the superscript ($c$) from $n_\r$ for simplicity.

The above deformation away from the pure Ising limit generates a complex energy landscape at low temperatures. Depending on the ratio, $V'/Va$ and $\ell/a$, the ground state can arrange into numerous ordered states at a sequence of commensurate fillings $n=p/q$ \cite{NovikovLevitov, Smerald_2016}. These ordered states range from various ``stripe'', or, ``Wigner-crystalline'' {\footnote{\textsf{The Wigner crystals are locked to the underlying lattice, where the periodicity is tied to the lattice constant.}}} patterns; the precise nature of these states will be discussed in Sec.~\ref{sec:thermo}. A schematic phase-diagram for the resulting model appears in Fig.~\ref{fig:fig1} (b). As a result of the longer-range interactions, the spectrum of the Hamiltonian gets modified due to the charge rearrangements that can arise over long distances. The massive ground-state degeneracy associated with the purely nearest-neighbor model introduced earlier is thereby lifted by an amount that is controlled by the strength of the longer-ranged interactions. The effect of the single-particle hopping can then be included simply within conventional perturbation theory \cite{Hartnoll18}, as we describe in Sec.~\ref{sec:transport} below. For the electronic model, the transport properties are controlled by nearest-neighbor electron hops, as long as the site occupancy allows for such hopping processes. Therefore, the conductivity at leading order in the hopping strength ($\sim O(t^2)$) is ultimately controlled by the density-density correlation functions in the corresponding classical model with $t=0$. As we shall discuss below, the broadening of a ``Drude-like'' peak in the optical conductivity is controlled by $V^{\prime}$.

\captionsetup[figure]{justification=centerlast}

\section{\textsf{Thermodynamics}}
\label{sec:thermo}

Let us now study the thermodynamic properties of the purely classical model with $t=0$. The key quantities of interest are the thermodynamic susceptibilities {\footnote{\textsf{These susceptibilities will play an important role in the full thermoelectric response, as discussed in appendix \ref{ap:TE}.}}, which are defined in the usual way
\begin{subequations}
\beq
\chi &=& -e^2 ~\frac{\partial^2 f}{\partial \mu^2},~~\zeta = -e~\frac{\partial^2 f}{\partial T\partial \mu},\\
c_\mu &=& -T \frac{\partial^2 f}{\partial T^2},~~c_{\rho} = c_\mu -\frac{T \zeta ^2}{\chi}, 
\eeq
\label{eqn:suscept}
\end{subequations}
where $f=-\ln Z/(\beta\times\tn{vol})$ ($\tn{vol}\equiv$ volume) is the free energy density. In the above definitions, $\chi$ is the charge compressibility, $\zeta$ is a mixed susceptibility, $c_\mu$ is the specific heat and $c_\rho$ is the heat capacity at fixed charge density.  

We begin by analyzing the charge compressibility as a function of filling and temperature in the classical model with $t=0$ in order to map out the full phase-diagram. As discussed previously, we will focus on the regime where the effects of $t$ can be studied perturbatively (i.e. for $\lambda\gg1$). Therefore, the results are strictly valid as long as $T\gg t$, but with no further restriction on $T/V$ as long as $V'$ is finite. In this limit, the set of occupation numbers $\{n\}$ completely define the eigenstates with energy, $E_{\{n\}}$, where
\beq
E_{\{n\}} &=& \frac{1}{2}\sum_\r  \ve_\r n_\r,~\tn{with} \label{En}\\
\ve_\r &=& V \sum_{\r'\in \tn{n.n.}~\r} n_{\r'} + V' \sum_{\r'\neq \r}\frac{\exp{\left( -|\r-\r'|/\ell\right) }}{|\r-\r'|} n_{\r'}.
\label{eqn:loc_ene}
\eeq

The typical set of configurations, $\{n\}$ where $n_\r\in  \{ 0,1 \}$, are generated using Monte Carlo (MC) simulations for a given chemical potential and temperature. Unless stated otherwise, we focus on the system size of 30$\times$30 in the remainder of this paper. The details of our MC simulations, including the finite-size effects, are provided in  Appendix~\ref{ap:MC}. The effective Hamiltonian and the partition function are then described by, 
\beq
H_\tn{eff} &=& E_{\{n\}} - \mu N_{\{n\}}, \\
Z &=& \sum_{\{n\}} e^{-\beta H_\tn{eff}},
\label{Heff}
\eeq
where $N_{\{n\}}=\sum_\r n_\r$.

\captionsetup[figure]{justification=centerlast}
 \begin{figure*}[t!]
    \centering
    \begin{subfigure}[t]{0.5\textwidth}
        \centering
        \includegraphics[width=1.05\linewidth,left]{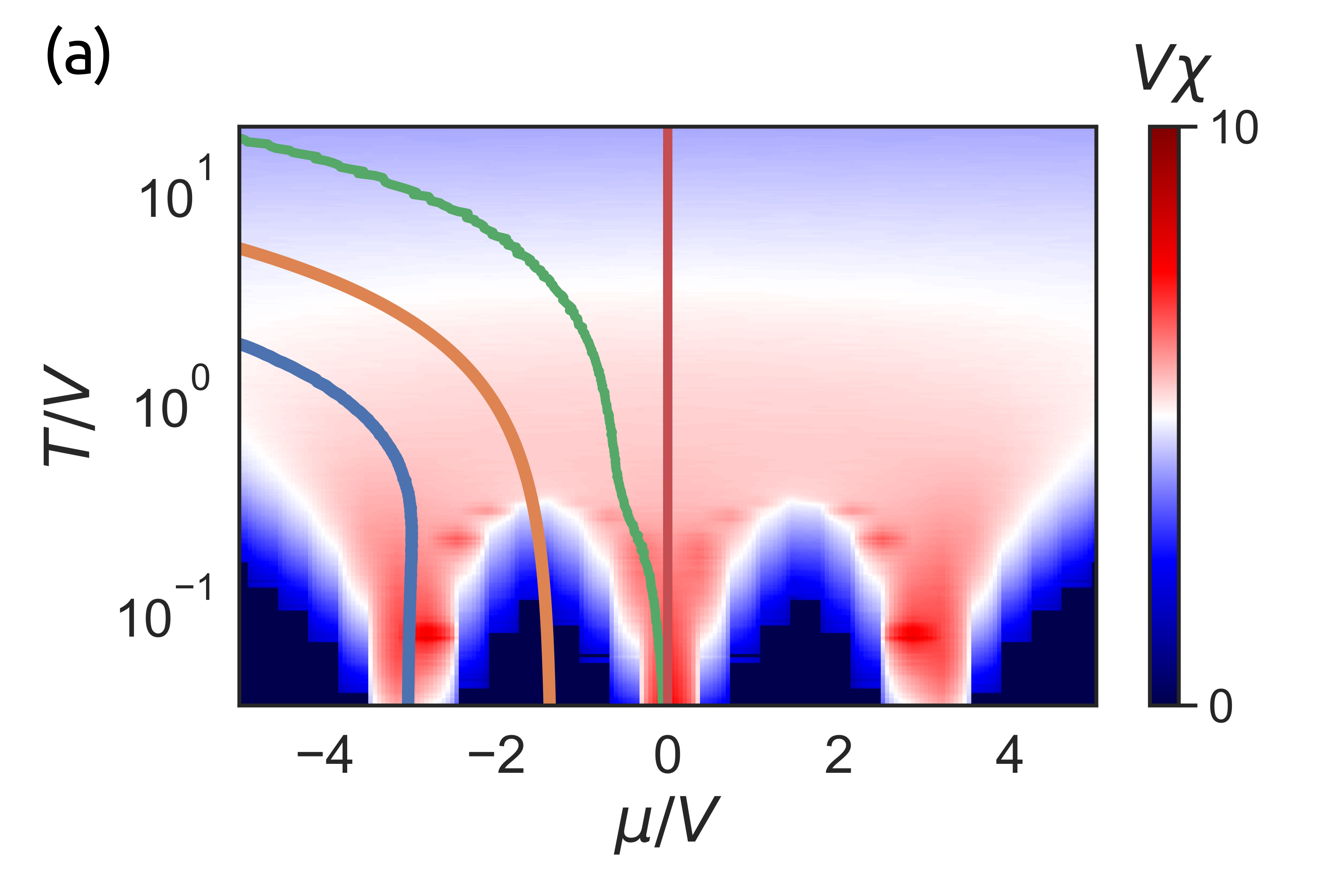}
    \end{subfigure}%
    ~ 
    \begin{subfigure}[t]{0.5\textwidth}
        \centering
        \includegraphics[width=1\linewidth, left]{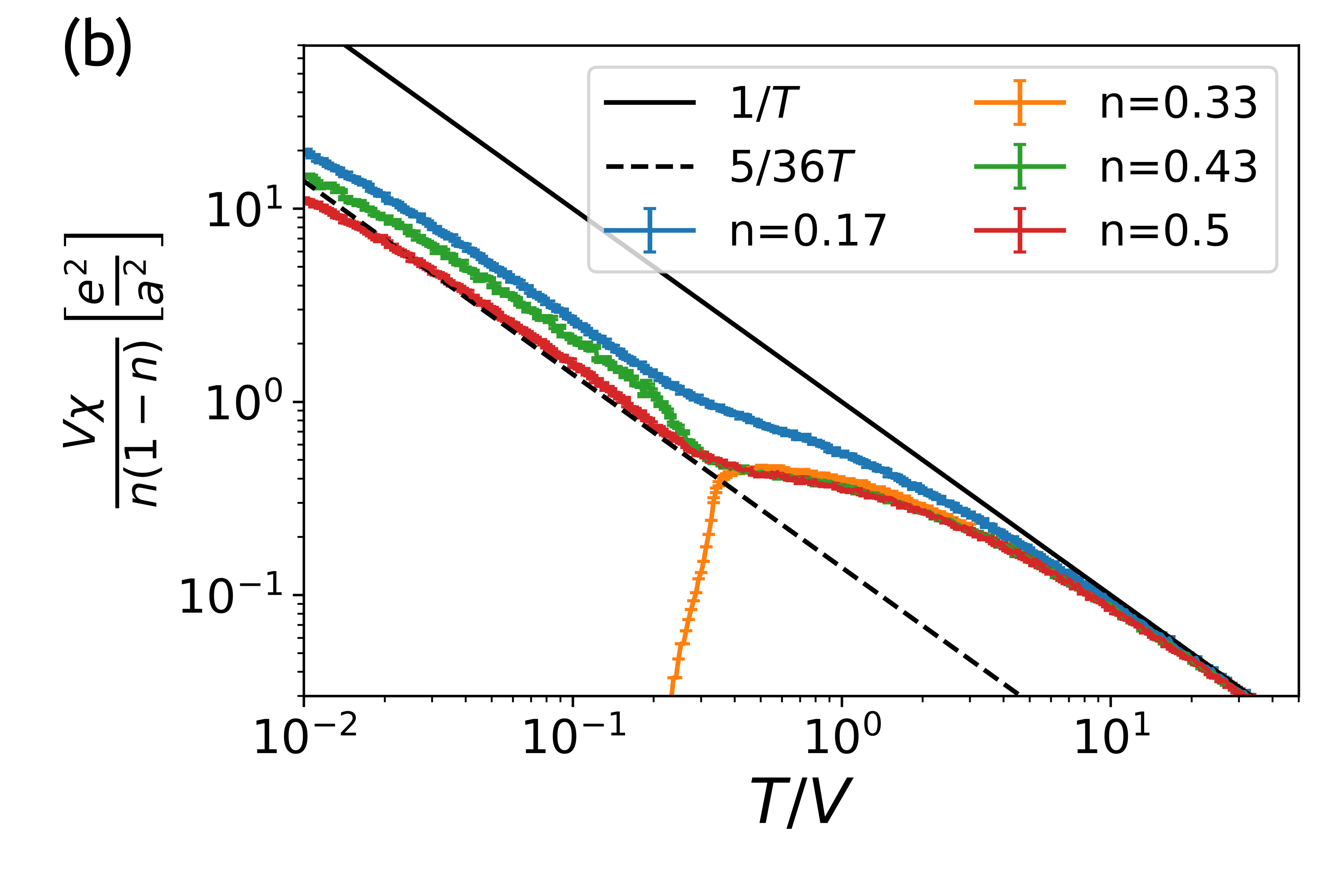}
    \end{subfigure}
    \caption{\textsf{(a) Compressibility as a function of temperature for $V'=0.01V$ for a range of chemical potentials. In the limit of low temperatures, four different incompressible states appear as domes over an extended range of $\mu$. The cooling curves corresponding to a fixed average density are shown as colored lines. (b) The rescaled compressibility calculated from the MC simulations shows distinct regimes of a $1/T$ behavior at high and low temperatures connected by a smooth crossover. (Error bars correspond to a $1\sigma$ standard deviation extracted from a bootstrap resampling; see Appendix \ref{ap:MC} for details). The solid and dashed lines correspond to asymptotic expansions for $\chi$ at high and low temperatures, respectively.}}
    \label{fig:phaseDiag} 
\end{figure*}

Using the fluctuation-response relation, we can express the compressibility in terms of the density fluctuations as,
\beq
\chi=\frac{e^2 \beta}{ \tn{vol} }\left(\left\langle N_{\{n\}}^{2}\right\rangle -\left\langle N_{\{n\}}\right\rangle ^{2}\right).
\label{eq:flucreschi}
\eeq
We begin by mapping out the compressibility in the entire temperature-filling phase diagram. At low temperatures and for the particular choice of $V'/(Va)=0.01$,   
{\footnote{\textsf{We set the lattice constant, $a=1$, from now onwards unless stated otherwise.}}} we find four distinct regions of the phase diagram with a vanishing compressibility. These regions, at fillings $n^*=0,~1/3,~2/3,~1$, are clearly visible as `dome-shaped' regions in Fig.~\ref{fig:phaseDiag}(a) with $\chi\rightarrow0$. Note that as a result of the particle-hole symmetric nature of the model (with $t=0$), all incompressible states come in pairs at fillings $n^*(=p/q)$, and, $1-n^*$, respectively.

For our Monte Carlo computations, in order to avoid freezing into false local minimum energy configurations, we initialize the simulations at high temperatures, $T \gg V$, and then cool the system adiabatically, allowing the system to sample larger regions in phase space (see Appendix \ref{ap:MC} for details). To fix the average density, $n= \langle N_{\{ n\} }\rangle /\tn{vol}$, we obtain the chemical potential needed to fix the average density using a binary search algorithm during the cool-down and letting the system thermalize at each step of the binary search \cite{Binarysearch}. The constant filling traces of the compressibility are shown as colored lines in Fig \ref{fig:phaseDiag}(a).

When considering the connection between our electronic model and the AFM Ising model with $t=0$, it is useful to note that the chemical potential in the former plays a role analogous to an external (longitudinal) magnetic field in the latter. Thus, the magnetic susceptibility in the AFM Ising model acts as the charge compressibility. The resulting analogy is especially useful for insight into the behavior of the charge compressibility and the onset of insulating (incompressible) behavior in the regime $t\ll T\ll V$, as shown in Fig.~\ref{fig:phaseDiag}(b). 

At high temperatures ($T\gg V$), we find that the chemical potential has the following asymptotic behaviour $\mu \rightarrow -T\log \left( 1/n -1\right)$, which leads to
\begin{equation}
    \chi (n,T) \rightarrow \frac{e^2}{a^2} \frac{n(1-n)}{T}. 
    \label{eqn:HighTcompressibility}
\end{equation}
The results from our explicit calculation using Eqn.~\ref{eq:flucreschi} at high temperatures are consistent with the above form. Moreover, the above result can be obtained from a high-temperature series expansion for generic models with a finite bandwidth at leading order in $1/T$ \cite{ChaikinThermopower, HubbHighTGeorges}; the density dependence is a consequence of the particle-hole symmetric form of the interaction-only model. A snapshot of a typical Monte Carlo configuration, $\{n\}$, in the high-temperature regime is shown in Fig.~\ref{fig:configs}(a) and reveals a thermally disordered state without any long-range ``liquid-like'' correlations. 

Let us now focus on the incompressible states in the limit of low temperatures. As is well known, at $n=1/3$, one obtains a $\sqrt{3}\times\sqrt{3}$ charge-density wave (CDW) ordered state as shown in Fig.~\ref{fig:configs}(c). A similar CDW (related by a particle-hole transformation) is the ground state at $n=2/3$. In the absence of further neighbor interactions, these are the only incompressible states at $T=0$ (apart from the trivial ones at $n=0,~1$). However, for a finite and sufficiently large $V'>0$, additional incompressible states can arise at various commensurate fillings; the corresponding transition temperatures to these states are then approximately determined by the the strength of beyond nearest-neighbor interactions \cite{NovikovLevitov, Smerald_2016}. For instance, when $V'/V\sim5$, we observe a freezing transition into a conmensurate CDW at $n=1/2$, as depicted in Fig.~ \ref{fig:configs}(d). The pattern shown in the figure corresponds to a stripe phase with three possible orientations \cite{StripePhaseTriag}. {\footnote{\textsf{For a sufficiently large $V'/V$, that favors freezing at other commensurate fillings, the complex energy landscape also leads to a slowing down of the Monte Carlo dynamics, which can result in the system freezing in a state with pockets of ordered regions separated by phase boundaries.}}} 

Our main interest here is in the transport properties at intermediate temperatures in the regime with liquid-like correlations (i.e. above any ordering tendencies). Therefore, we intentionally focus on the situation where $V'/V$ is small enough such that there is no freezing transition into any of the crystalline states except for at $n=1/3,~2/3$. Moreover, this also provides us with a broad intermediate range of temperatures where the transport properties can be analyzed reliably and interesting connections can be made by appealing to the proximity to the Ising AFM phenomenology at $V'=0$.

\captionsetup[figure]{justification=centerlast}
\begin{figure*}
        \centering
        \begin{subfigure}{0.45\textwidth}
            \centering
            \includegraphics[width=1\linewidth]{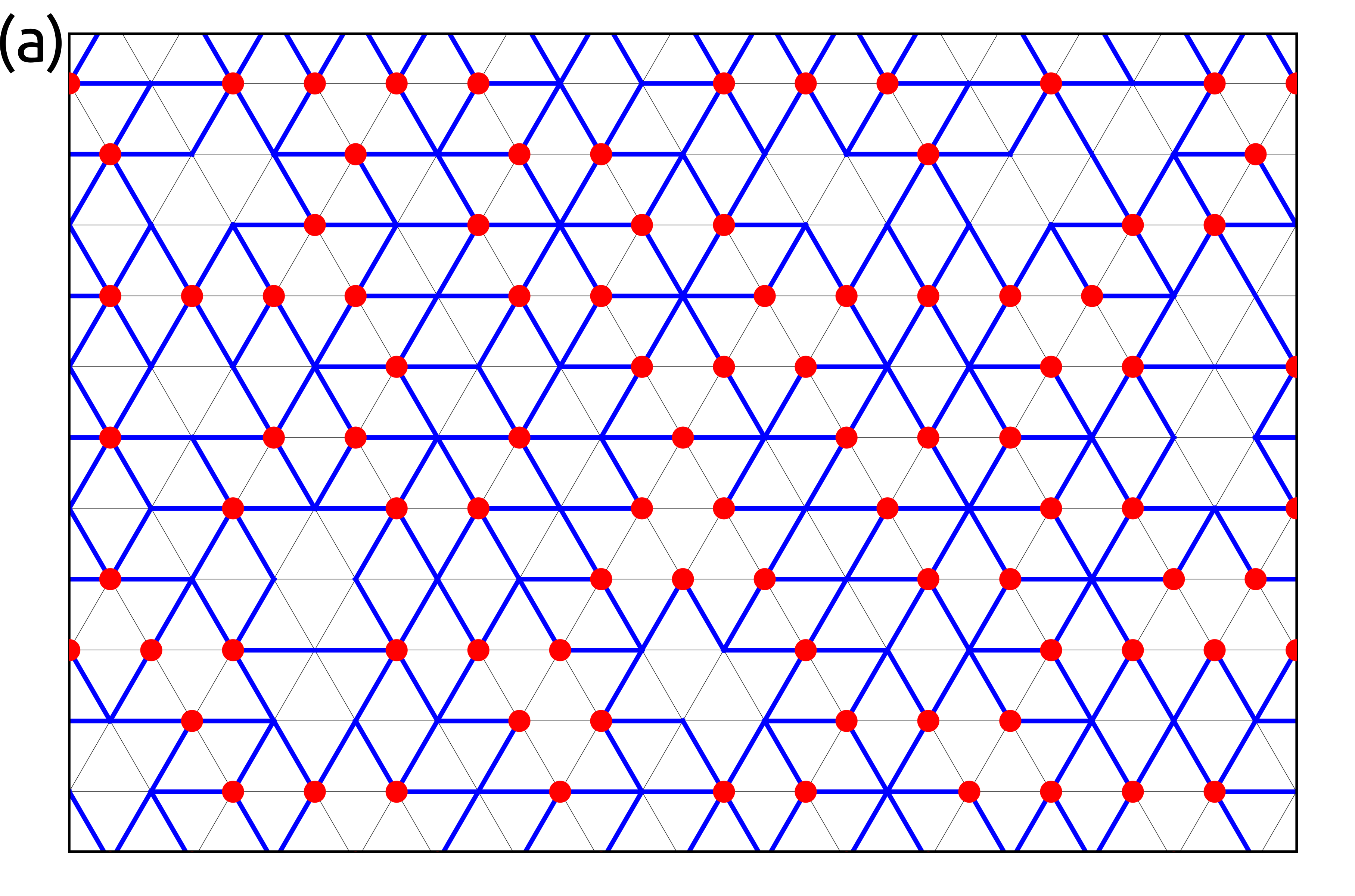}
            
        \end{subfigure}
        ~
        \begin{subfigure}{0.45\textwidth}  
            \centering 
            \includegraphics[width=1\linewidth]{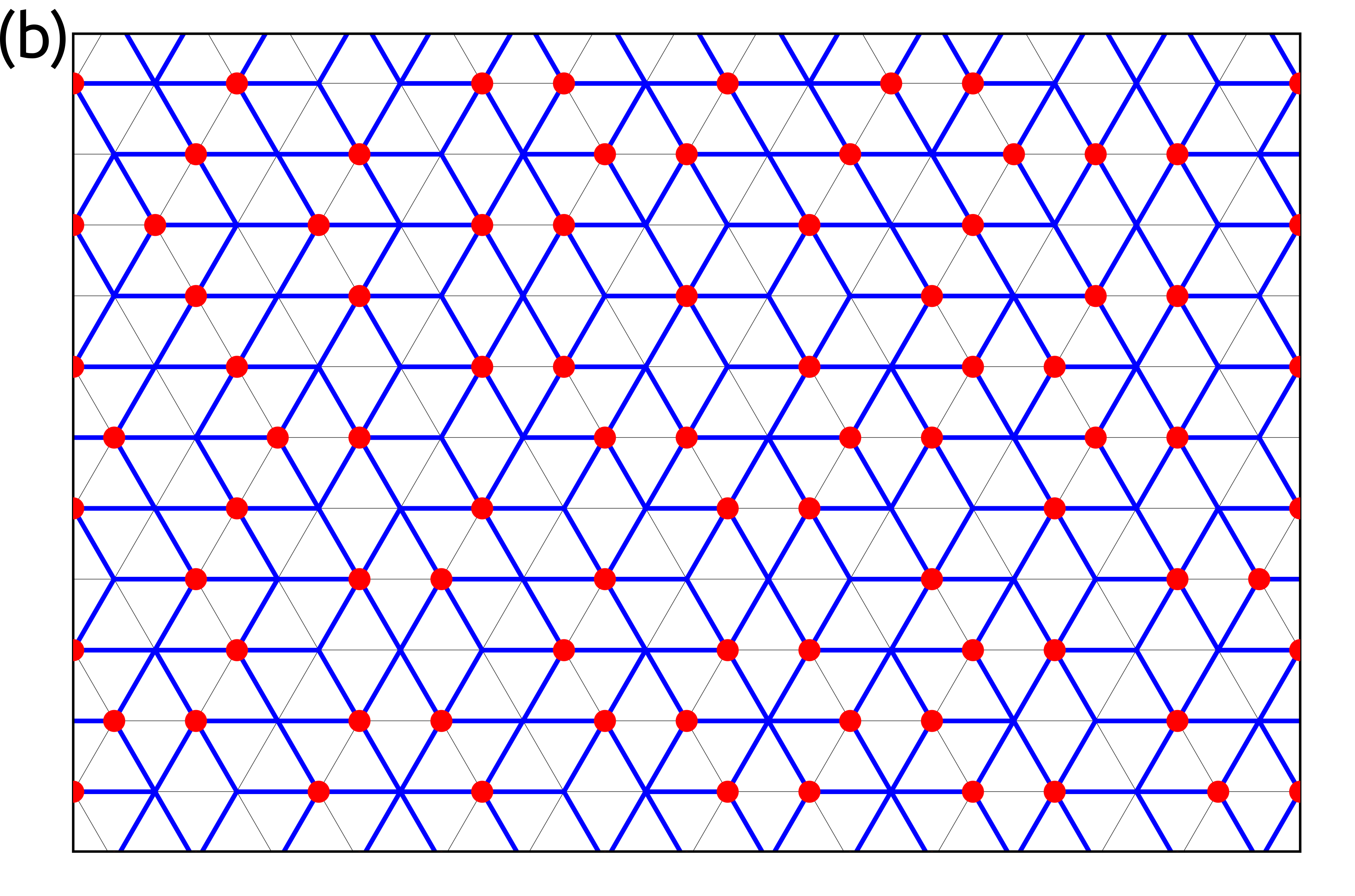}
          
        \end{subfigure}
        \vskip\baselineskip
        \begin{subfigure}{0.45\textwidth}   
            \centering 
            \includegraphics[width=1\linewidth]{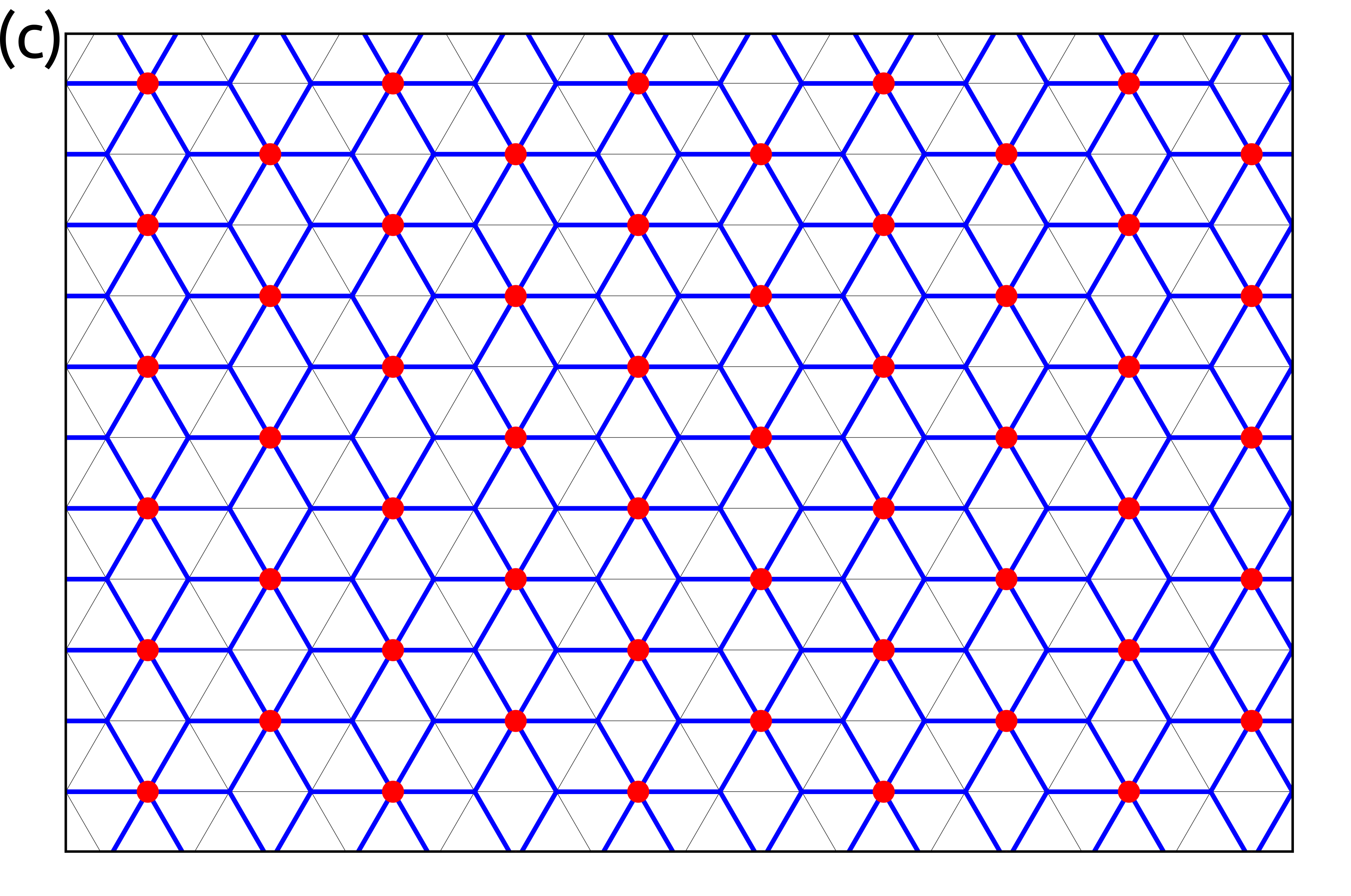}
            
        \end{subfigure}
        ~
        \begin{subfigure}{0.45\textwidth}   
            \centering 
            \includegraphics[width=1\linewidth]{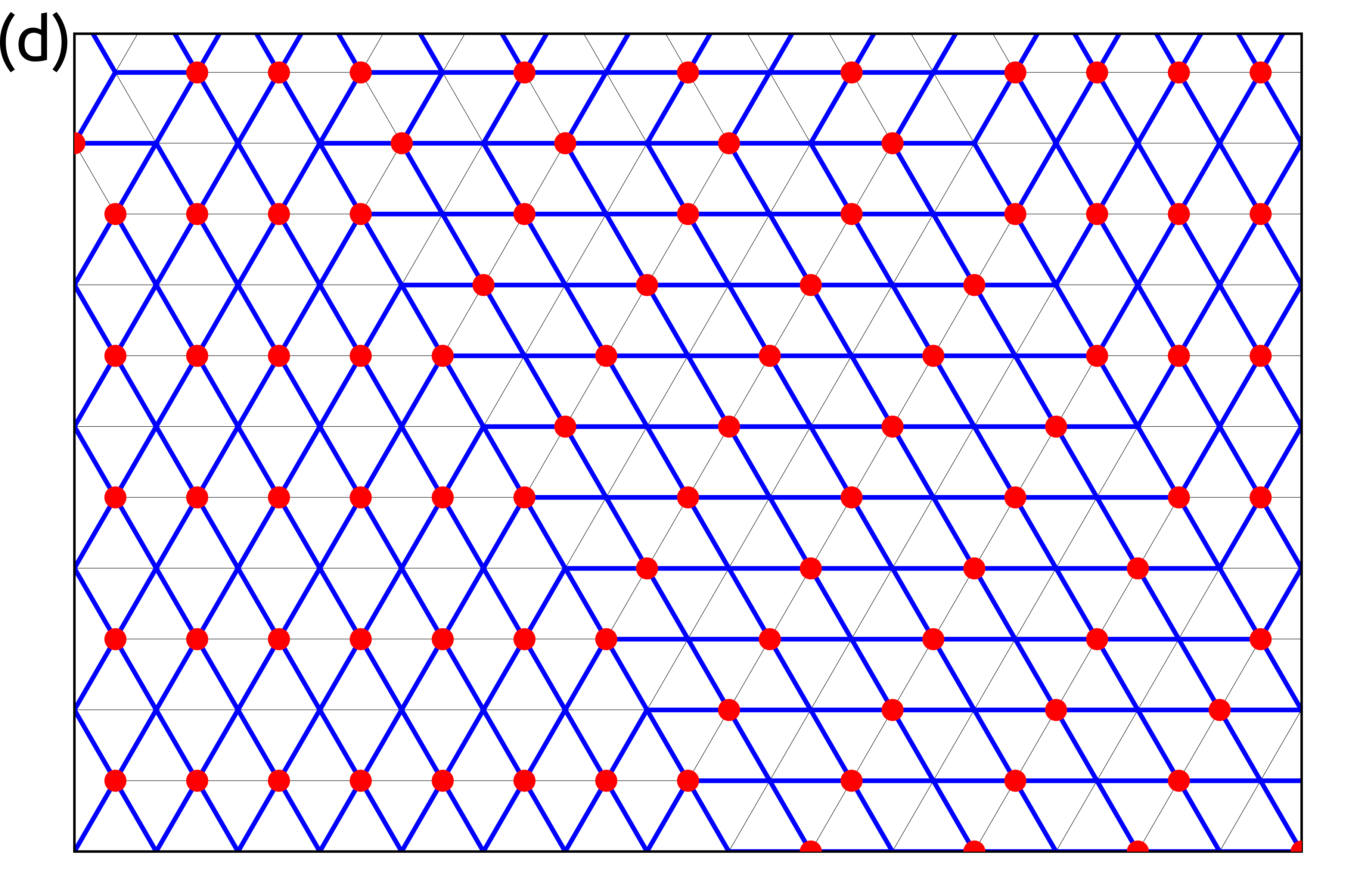}
        \end{subfigure}
        
        \caption{\textsf{Typical configurations generated in our Monte-Carlo simulations at different temperatures. Red dots represent occupied sites and blue lines denote the unfrustrated bonds (see text for discussion). (a) At half-filling and for $T\gg V$, the charge configuration is uncorrelated without any restriction on the number of frustrated bonds in a triangle. (b) A snapshot of a regime with liquid-like, long-range correlations at $T=0.1 V$. The configurations are in a one-to-one correspondence with rhombic tillings of the plane; the defects in the tilling are exponentially suppressed at low temperatures. (c) Charge ordered state at $n=1/3$ at $T=0.09V$. (d) Stripe phase at half-filling determined by the strength of further neighbor interactions at $T=0.01V$ with $V'=5V$.}}
        \label{fig:configs}%
    \end{figure*}

We have noticed that for a range of generic fillings (i.e. away from special commensurate fillings), the charge compressibility continues to have the same $\chi\sim1/T$ temperature dependence at high $T\gg V$ as well as at low $T\ll V$. It is natural to ask if the temperature independent prefactors in these two distinct regimes are the same or different. This question can be addressed analytically at half filling when $t=0$ by returning to the language of Ising spins and by recalling that the susceptibility is related to the compressibility in the electronic language. Quite remarkably, the susceptibility of the AFM Ising model on the triangular lattice can be related to the sum of the susceptibilities of a FM and an AFM Ising model defined on a honeycomb lattice \cite{Fisher59}. At half-filling, this leads to a low-temperature form: $\lim_{T\rightarrow 0}~T\chi=B\lim_{T\rightarrow \infty}T\chi$ with $B=5/36$ \cite{antiferrosuceptHF}. Our numerical results agree with this result (see Fig.~\ref{fig:phaseDiag} b) and the two distinct regimes of $1/T-$compressibility are connected by a smooth crossover at a temperature $T\sim V$.  Numerically, we have observed similar qualitative behavior for a range of other fillings but with different slopes in the low-temperature regime (except at the various commensurate fillings where the liquid freezes into an ordered state, as stated earlier), as shown in Fig.~\ref{fig:phaseDiag} b. Thus, the inherent geometric frustration in the model prevents any ordering tendencies in the parameter range under study. 

The intermediate regime with liquid-like correlations is most clearly obtained in the temperature window $V'<T<V$ and for fillings $1/3<\nu<2/3$. A snapshot of the typical Monte Carlo configuration, $\{n\}$, in this regime is as shown in Fig.~\ref{fig:configs}b. The compressibility in this regime is finite and the liquid state is characterized by the approximate constraint of having only one frustrated bond per triangle \cite{NovikovLevitov,chamon2017topological}. At finite temperatures, the violation of the constraint by thermal fluctuations is possible, but with an exponentially small probability and states that violate this constraint are almost completely absent in the MC simulation.

\captionsetup[figure]{justification=centerlast}
\begin{figure}[ht!]
    \centering
    \includegraphics[width=14cm]{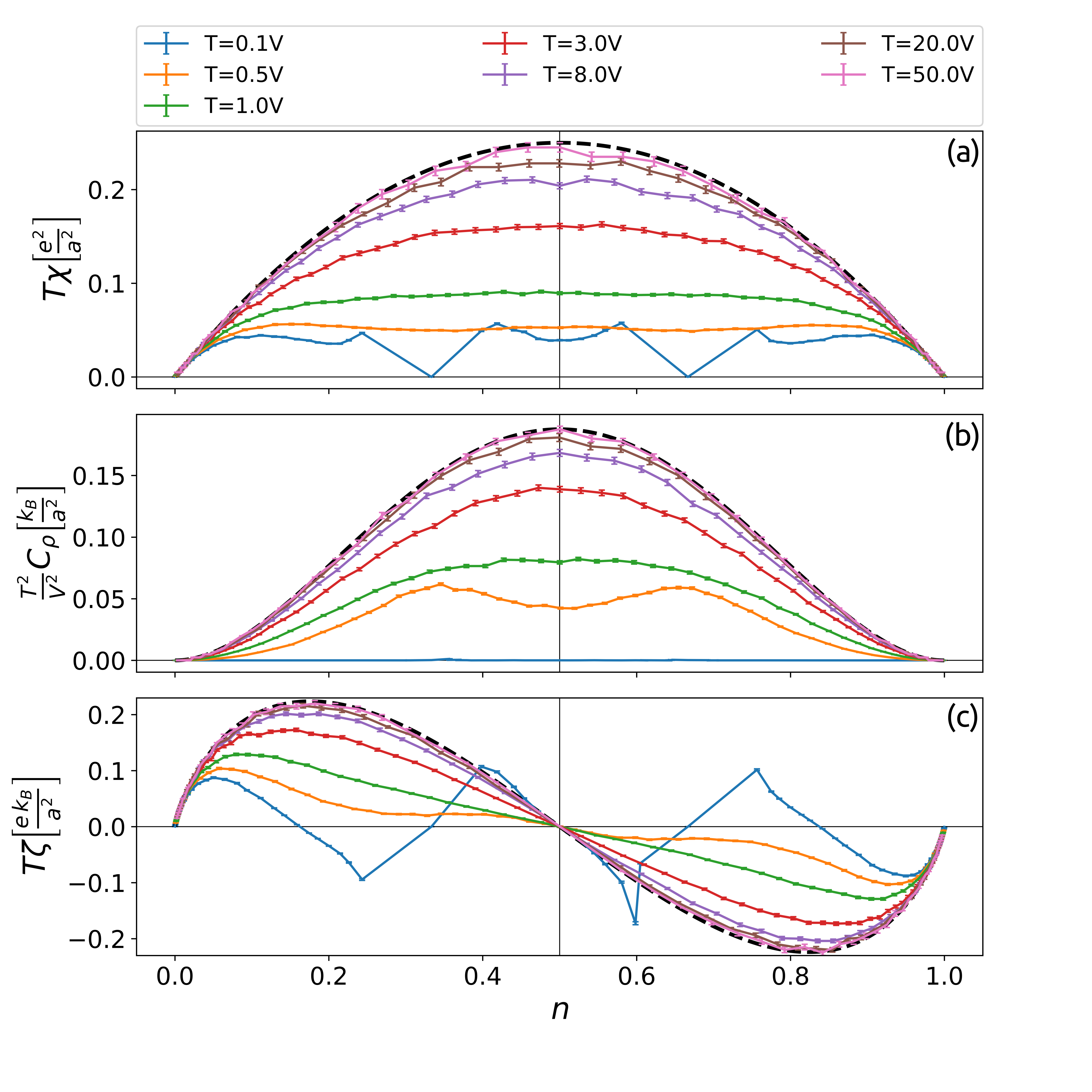}
    \caption{\textsf{Thermodynamic susceptibilities at fixed temperatures as a function of filling. The susceptibilities have been re-scaled by appropriate factors of $T$ such that their high temperature limit is $T-$independent. The dashed line in each figure corresponds to the result expected from the high-temperature expansion.}}
    \label{fig:ThermoQ}%
\end{figure}

We now focus on the crossover from the high-temperature regime to the above intermediate regime. In particular, we study the thermodynamic quantities introduced earlier in Eqn.~\ref{eqn:suscept} (in addition to charge compressibility) as a function of filling. Starting with the high temperature limit where $\mu\propto T$, the range of $\mu$ has to be adjusted carefully to sweep through the values of density that are of interest to us as we cool the system. \footnote{\textsf{In this procedure, we trade off the adiabatic cooling in our algorithm in exchange for the detailed determination of the $n$ dependence at intermediate $T$. Note that by omitting adiabatic cooling in the procedure above, the results are reliable for fillings away from $1/3, 2/3$ at low temperatures.}} The results for $\chi,~C_\rho$ and $\zeta$ are shown in Fig. \ref{fig:ThermoQ} over a broad range of temperatures and fillings, $0\leq n\leq1$. In order to gain some insight, we first compare these results with a standard high-temperature expansion of the associated thermodynamic functions at $T\gg V$. In this limit, the compressibility takes the form as in Eq.~\ref{eqn:HighTcompressibility}. The leading order contributions to the other susceptibilities are given by, 
\begin{subequations}
\beq
T \zeta & \sim& \frac{k_B e}{a^2} n (1-n) \log \frac{n}{1-n},\\
T^2 c_\rho &\sim& \frac{k_B}{a^2} n^2 (1-n)^2.
\eeq
\label{eqn:susceptHighT}
\end{subequations}
It is important to recall that the leading order contribution to $c_\rho$ at $O(\beta)$ vanishes as a result of an exact cancellation between $c_\mu$ and $T\zeta^2/\chi$. For the given system-size ($30\times30$), we observe quantitative agreement between our simulations and the result for the high-temperature expansion at $T\gg V$. At intermediate fillings, there is a systematic departure between the two sets of results upon cooling even at $T\gtrsim V$ (see Fig.~\ref{fig:ThermoQ}), which is unrelated to any finite-size effects; the magnitude of the deviation is likely controlled by the next to leading order corrections in the series expansion. We also note in passing that the ratio $\zeta /\chi$, which is related to the thermopower, is controlled by the high-temperature expansion of $\mu/T$. As a result, the asymmetry of $\zeta$ is controlled by the particle-hole asymmetry, whose overall amplitude is controlled by the compressibility.

As a function of decreasing temperature, all of the susceptibilities start to develop pronounced features and large fluctuations near the commensurate fillings $n=1/3,~2/3$ at approximately $T\approx 0.1V$; see Fig.~\ref{fig:ThermoQ}.\footnote{\textsf{The further neighbor interactions, $V'(\ll V)$, leads to the onset of similar features at other commensurate fillings that are not of particular interest to us in this paper.}} As was already pointed out in Fig.~\ref{fig:phaseDiag}b, the compressibility falls precipitously roughly at the same temperature at these fillings, signaling the transition into an insulating state. On the other hand, away from these insulating fillings the compressibility is weakly filling dependent with a $\chi\sim 1/T$ dependence on the temperature, as in Fig.~\ref{fig:phaseDiag}b. There is thus a crossover in the compressibility from a high-temperature $1/T$ behavior to an intermediate temperature $1/T$ regime, where the latter is controlled by the underlying frustration of the triangular lattice model. The heat capacity shows a dip at intermediate fillings (Fig.~\ref{fig:ThermoQ} b) before falling abruptly at low temperatures; see Appendix \ref{ap:Is} for the low temperature behavior of $c_\rho$. The dip at intermediate fillings is related to the reduction of energy fluctuations upon entering the correlated fluid phase. Finally, $\zeta$ undergoes multiple sign changes at low temperatures, in contrast to a single sign-changing behavior across half-filling at high temperatures. This is a hallmark of the sign change associated with the nature of the effective carriers at these low temperatures. At high temperature ($T\gg V$), the charge transport in the uncorrelated fluid can be interpreted simply in terms of particles and holes (relative to $n=0$ and $n=1$) depending on whether $n<1/2$ or $n>1/2$.  However, the situation is complex at low temperatures as a result of the freezing transitions at commensurate fillings. Thus transport in this regime is dictated by the nature of ``doped'' charge carriers relative to these insulators, leading to the additional sign-changes in $\zeta$.

\section{Transport}
\label{sec:transport}

\subsection{Optical Conductivity}
\label{sec:opt}

Let us begin by computing the charge-transport properties within linear response theory using Kubo formula. The longitudinal conductivity, let us say along the $x-$direction, can be calculated as
\beq
\sigma^{xx}(\omega) \equiv \frac{1}{\tn{vol}} \int_0^\infty d\tau~e^{i\omega^+\tau} \int_0^\beta d\lambda \langle J^x(\tau-i\lambda) J^x\rangle,
\label{cond}
\eeq
where the thermal expectation value is evaluated using the effective Hamiltonian in Eq.~\ref{Heff}: $\langle...\rangle = \tn{Tr}(...e^{-\beta H_{\tn{eff}}}/Z)$. Note that since the electron current operator $|\vec{J}|\propto t$, in the perturbative regime of interest with $\lambda\gg1$, we can evaluate the thermal expectation values in the purely classical theory with $t=0$. Therefore, we also immediately obtain that the charge conductivity, $\sigma^{xx}\propto t^2$. The temperature and interaction strength dependence of the conductivity then arises solely from the correlators of the density operators for nearest-neighbor sites (these are the sites connected by the nearest-neighbor electron hops) in the theory described by $H_\tn{eff}$.  The conductivity at a finite frequency, $\omega$ is then given by, 

\beq
\sigma^{xx}\left(\omega\right)=\frac{e^{2}}{h} t^2 \frac{1-e^{-\beta\hbar\omega}}{\hbar \omega}\sum_{\{n\}}\frac{e^{-\beta H_{\tn{eff}}}}{Z}\frac{2\pi^{2}}{\tn{vol}}\sum_{\left\langle \r,\r'\right\rangle }\left( D_{\r\r'}^{x}\right)^{2}\Delta_{\r\r'}(\omega),\label{eqn:optcond}
\eeq 
where $D_{\r\r'}^{x}$ is the $x$ component of the vector $\bm{D}_{\r\r'}=\r - \r'$, and 
\begin{equation}
\Delta_{\r\r'}(\omega)=n_{\r^{\prime}} (1-n_{\r})~\delta(\hbar \omega - \varepsilon^{f}_{\r} + \varepsilon^{i}_{\r^{\prime}} )
\label{eqn:spec_weight}
\end{equation}
enumerates the number of excitations with energy $\hbar\omega$ that can be generated for a single electron hop between nearest neighbor sites $\r$ and $\r'$. The energy difference corresponds to moving an electron at $\r^\prime$ with local energy $\varepsilon_{\r^\prime}^{i}$ to a nearest neighbor site $\r$ with local energy $\varepsilon_{\r}^{f}$, where the energies are given by Eqn.~\ref{eqn:loc_ene}; see Appendix  \ref{ap:LR} for a detailed derivation. It can be readily verified that $\sigma^{xx} (\omega)$ is invariant under particle-hole transformations. We also note that the expectation value in Eqn.~\ref{eqn:optcond} effectively amounts to computing the histogram of $ \Delta  \varepsilon_{\r\r'}=\varepsilon^{f}_{\r} - \varepsilon^{i}_{\r^{\prime}}$. 

\begin{figure}
\centering
\includegraphics[width=1\linewidth]{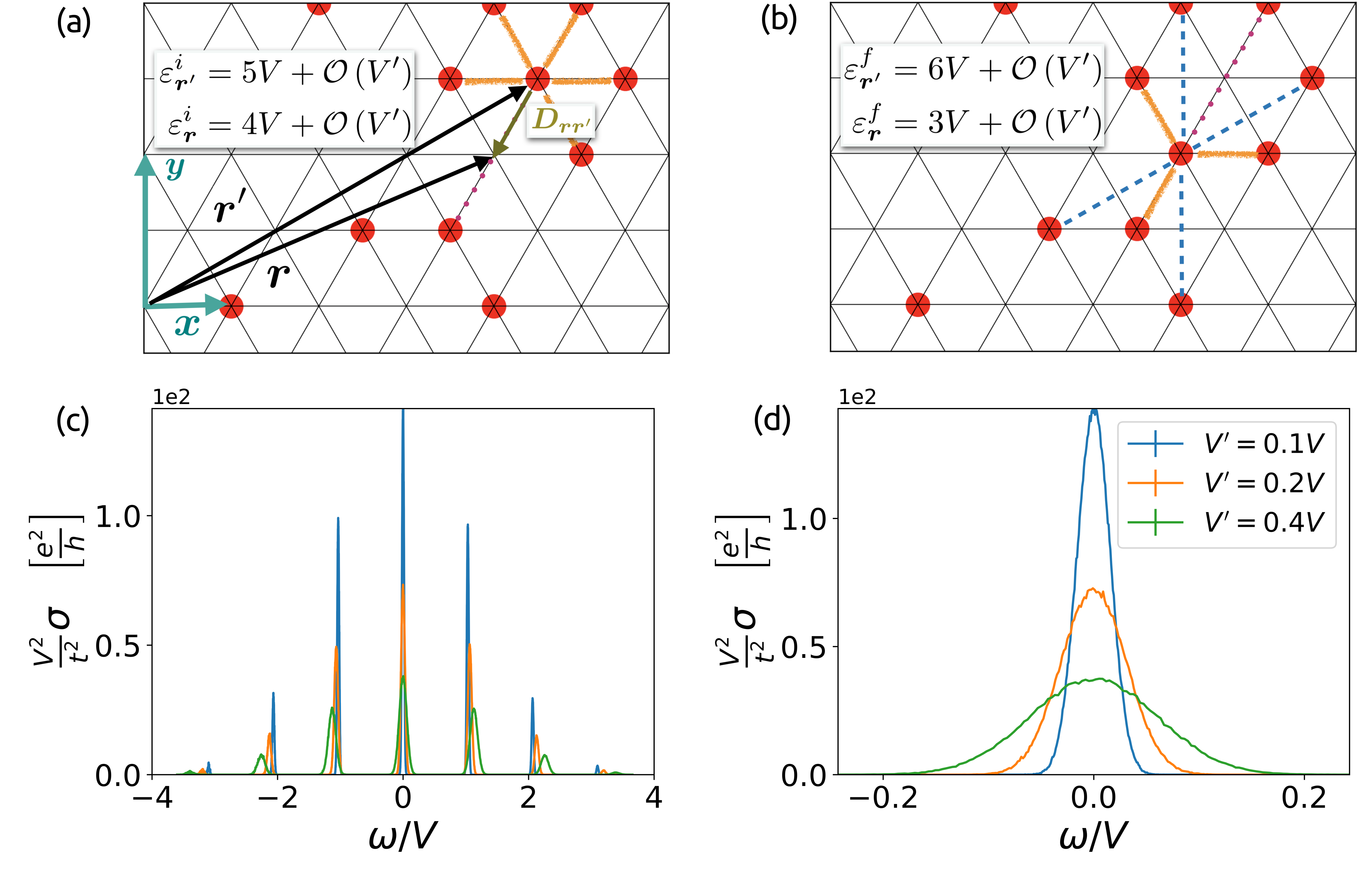}
\caption{\textsf{(a) and (b) Typical configurations associated with an electron hop  from $\r'$ to $\r$ and the corresponding energetics. Orange lines denote bonds that contribute to $\varepsilon_{\r^\prime}^{i}$. Dashed (dotted) lines represent second (third) nearest neighbour interactions, respectively. The local energies at leading order (i.e. ignoring contributions at $O(V')$) are summarized in the inset.
        The change in the local energy associated with the particular hopping process contributes to the spectral weight in the optical conductivity at a frequency $\sim 2V$. c) Optical conductivity calculated using Eqn.~\ref{eqn:optcond} at $T=V$ as a function of increasing $V'/V$. d) A magnified view of the central peak in (c).}}
        \label{fig:optfig}%
\end{figure}

Based on our discussion thus far, it would appear that the hopping strength only appears as a multiplicative factor ($\sim t^2$) in the conductivity and sets the overall scale of the charge response. However, there is an additional subtlety imposed by the particular hierarchy of energy scales that we have invoked in the problem. Formally, our approach is only valid at frequencies, $\hbar \omega > t$, and in order to incorporate the lack of an energy resolution at  $\hbar \omega \lesssim t$, we impose $t$ as a lower cutoff on the bin-size that we use in the histogram of $ \Delta  \varepsilon_{\r\r'}$. We choose $t<V^{\prime}$ and we have verified explicitly that our results for $\sigma^{xx}(\omega)$ do not depend sensitively on a particular numerical choice of $t$.

We show the results for $\sigma^{xx}(\omega)$ as a function of $\omega$ at half filling and a fixed $T=V$ in Fig.~\ref{fig:optfig} for several values of $V^{\prime}>0$. The first feature worth noting is the central peak with the largest magnitude, flanked by six ``satellite'' peaks on either side of $\omega=0$. This is not a coincidence and can be understood by taking into consideration the coordination number of the triangular lattice, and, the energetics associated with the hopping processes. Consider a single fermion hopping process from an occupied to an adjacent empty site. An example of such a process that contributes to the spectral weight, $\Delta_{\r\r'}(\omega)$, is displayed in Fig. \ref{fig:optfig}. The difference in the local energy of the fermion as it hops from $\r$ to $\r'$ does not receive any contribution from sites that are simultaneously nearest neighbours of $\r$ and $\r'$. For instance, in Fig.~\ref{fig:optfig}, these would be the sites labeled by $\r + \hat{\bm{x}}$ and $\r' - \hat{\bm{x}}$, which together contribute an energy $2V$ to $\varepsilon_\r ^f$ and $\varepsilon_{\r'}^i$. Thus, the only remaining contribution to $\Delta \varepsilon_{ \r \r'}$ arises from occupied sites that are not shared nearest neighbours between $\r$ and $\r'$. There are three such sites for both $\r$ and $\r'$. Therefore, the value of $\Delta \varepsilon_{\r \r'}$  will be approximately given by one of the possible values $\{0,V,2V,3V\}$, up to terms of order $V'$. Consequently, summing over the allowed configurations in order to obtain $\sigma ^{xx} (\omega)$ in Eqn.~\ref{eqn:optcond} leads to the largest contribution centered around  $\omega=-3V,\cdots,0 ,\cdots, 3V$.

In order to study the role of the extended interactions on the optical conductivity, we study the behavior of $\sigma^{xx}(\omega)$ as a function of $V'/V$. Increasing $V'/V$ has two clear effects, as we demonstrate in Fig.~\ref{fig:optfig} c,d. First of all, as a result of the change in the energy differences involved with the fermion hopping process, the locations of the six satellite peaks (i.e. all but the one centered at $\omega=0$) shift by an amount proportional to $V'$.  Secondly, the ratio  $V'/V$  controls the width of each of the peaks in the optical conductivity. With a decreasing $V'/V$, the peaks become narrower; in the asymptotic limit where $V'\rightarrow0$ (which lies beyond the scope of our analysis), the associated width for these peaks also tends to zero. Thus a finite $V'/V$ and the associated $\delta\Hc$ is responsible for generating a finite ``optical'' scattering rate within our approach. We note that in the absence of a finite $V'/V$ and away from the limit that we treat perturbatively in $t/V (=\lambda^{-1})$, it is likely that a finite optical scattering rate is generated in a non-perturbative fashion within the highly degenerate ground-state manifold. A treatment of this non-trivial regime lies beyond the scope of our present discussion. 

To conclude our discussion of the optical conductivity, we note that varying $V'/V$ can in principle lead to a dramatic spectral rearrangement and a transfer of spectral weight from one peak to another. However, this requires a sufficiently large value of $V'$, such that the $V'-$dependent $\Delta \varepsilon_{\r \r'}$ becomes comparable to $V$. In the regime $V \gg V'$, such a spectral rearrangement is absent. Moreover the f-sum rule \cite{Kubo_response} imposes a further constraint on the total integrated spectral weight in the total optical conductivity; as a result of an absence of any spectral rearrangement we have also observed that the integral within each peak is approximately separately conserved. Therefore, as the peaks broaden with an increasing $V'/V$, the height diminishes. As a result, if we were to focus on the central peak near $\omega=0$ at a fixed $T$, the optical scattering rate is proportional to $V'$ and the value of $\sigma^{xx}(\omega\rightarrow0)$ is inversely proportional to $V'$ in the limit $V \gg V'$.

\subsection{DC Resistivity}
\label{sec:dctransport}

Let us now turn our attention to the dc resistivity. As noted earlier, in the strong-coupling regime of interest to us ($\lambda\gg1$), the dc resistivity can not be described using conventional Boltzmann theory of transport. Instead, as we discuss below, the finite resistivity arises from local electron hopping processes and Umklapp scattering in the self-consistently generated complex landscape that is obtained using classical Monte Carlo calculations. Empirically, we find that the shape of the central peak in Fig.~\ref{fig:optfig} is well fit by a Gaussian, $\sigma (\omega) = \mathcal{D}\tau\exp\left(-\pi\omega^2 \tau^2\right)$, where $\tau$ is the current relaxation time and $\mathcal{D}$ is a proxy for a ``Drude-weight''. In order to extract the dc resistivity, $\rd$, we focus our attention on this central peak and extrapolate to the limit of $\omega\rightarrow0$. We repeat the procedure at different temperatures and thereby extract the temperature dependent dc resistivity; see Fig.~\ref{fig:Res}. Note that in what follows, we will not attempt to decompose $\rd$ in terms of a scattering rate and Drude-weight separately. A similar functional form for $\sigma(\omega)$ has been pointed out earlier in a variety of different settings \cite{Hartnoll18,lindner,Huang987}.

At high temperatures ($T\gg V$) and at half-filling, our results for $\rd$ can be understood in an especially straightforward fashion. The classical correlations, as obtained from the Monte-Carlo calculations, are completely thermally disordered (see Fig.~\ref{fig:configs}a). As is evident from the behavior of $\rd$ in the inset of Fig.~\ref{fig:Res}a, there is a broad range of temperatures over which the dc resistivity is perfectly linear (i.e. $\rd\sim T$). In a system with a finite bandwidth ($\propto t$), such a high-temperature regime is naturally expected to display $T-$linear resistivity \cite{Oganesyan,HubbHighTGeorges}. This follows directly from conditions of local equilibrium and the Einstein-Nernst relation. In general, thermoelectric effects can play an important role in determining coupled charge and energy diffusion. However, such effects are absent at half-filling, as also verified explicitly in Appendix \ref{ap:TE}. Thus, we can express the conductivity simply as $\sigma_{\tn{dc}} =D_c  \chi$, where $D_c$ is the charge diffusion coefficient and $\chi$ is the compressibility. In the high-temperature limit of interest, the compressibility $\chi\sim 1/T$ , as was already discussed in Fig.~\ref{fig:phaseDiag}b. {\footnote{\textsf{The high-temperature $1/T$ behavior of the compressibility survives for the full quantum mechanical model, irrespective of the value of $\lambda$.}}} On the other hand, $D_c$ is expected to be largely temperature independent (i.e. the corresponding transport ``scattering-rate'' becomes nearly temperature independent). Thus, the $T-$linear behavior of the resistivity is primarily due to the temperature dependence of the compressibility (as in Eqn.~\ref{eqn:HighTcompressibility}). Even at densities away from half-filling, the slope $d\rd/dT$ is approximately independent of the density in the $T\sim V$ regime (see Fig.~\ref{fig:Res}) and the thermo-electric coefficients can be neglected; see Appendix \ref{ap:TE}. As a result, we are immediately led to the conclusion that the diffusion coefficient, $D_c^{-1}\sim n(1-n)$ for $T\gg V$ to offset the density dependence of the compressibility. 

\captionsetup[figure]{justification=centerlast}
\begin{figure*}[t!]
    \centering
    \begin{subfigure}[t]{0.5\textwidth}
        \centering
        \includegraphics[width=0.95\linewidth]{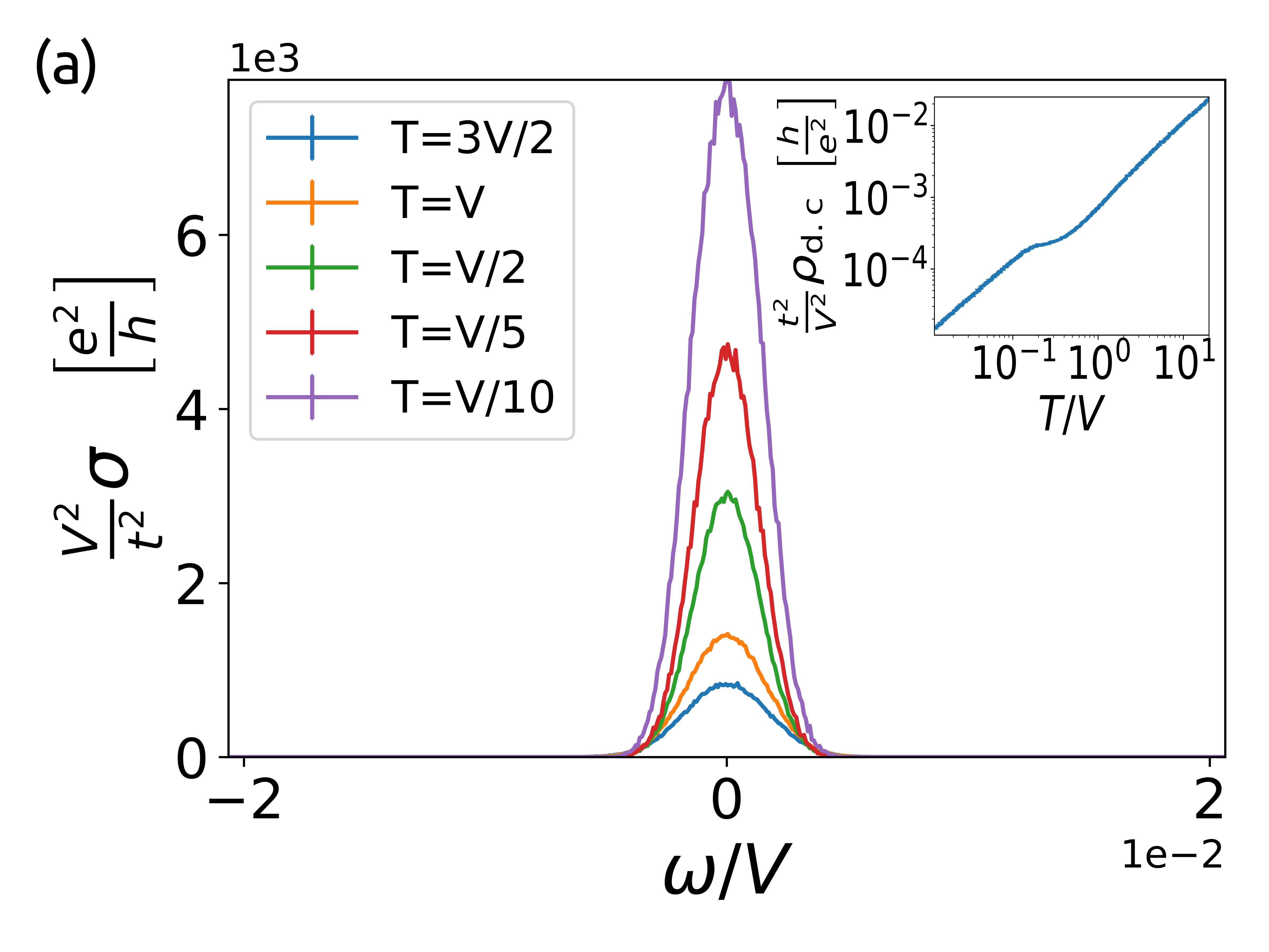}
    \end{subfigure}%
    ~ 
    \begin{subfigure}[t]{0.5\textwidth}
        \centering
        \includegraphics[width=1.05\linewidth]{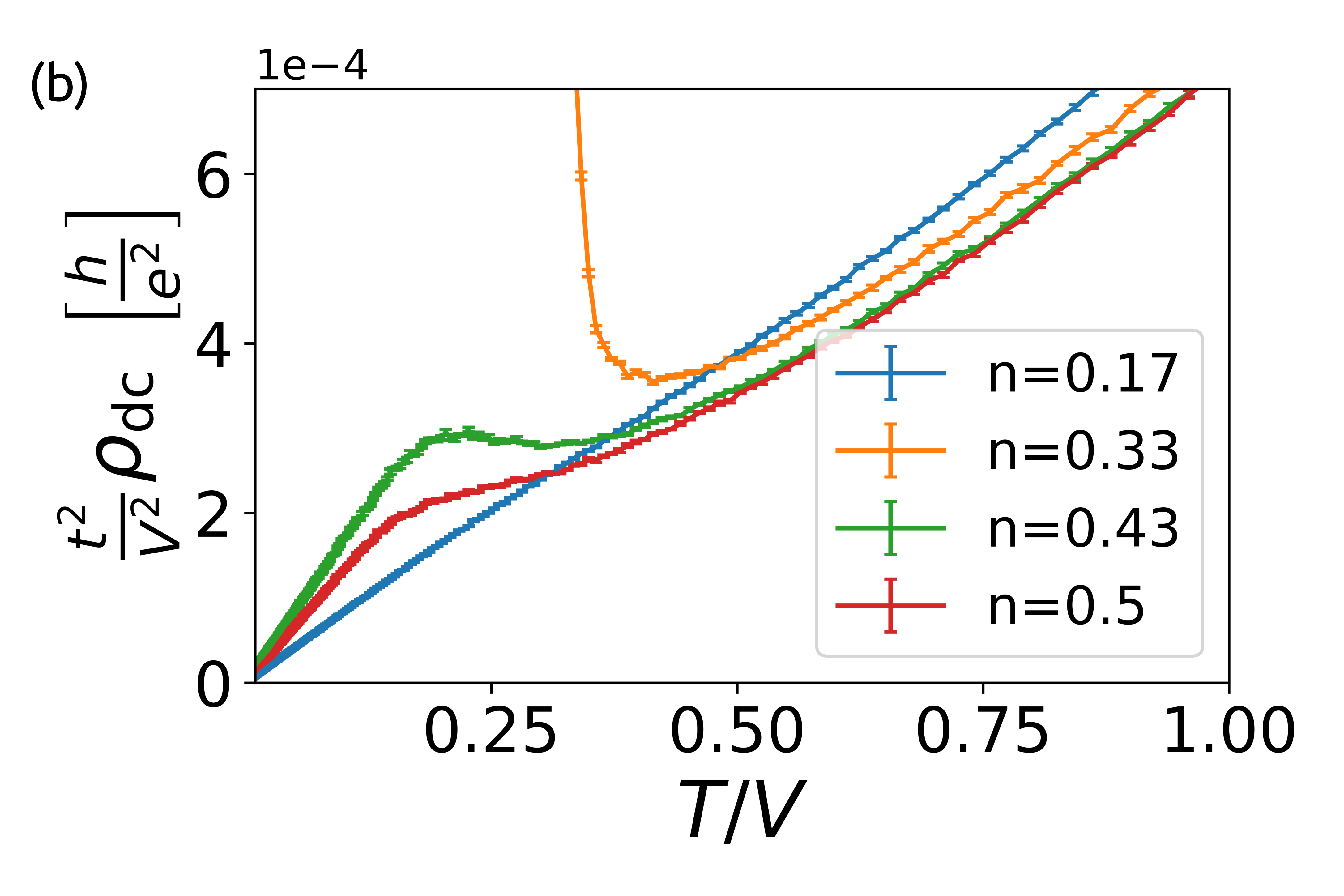}
    \end{subfigure}
    \caption{\textsf{a) The central peak of the optical conductivity for fixed $V^{\prime}=0.01V$, $t=0.1 V^{\prime}$ and $n=1/2$ as a function of decreasing temperature. The dc resistivity is extracted from the peak-height, which is strongly temperature dependent; see inset. The peak-width is mildly temperature dependent. b) The dc resistivity as a function of temperature for a few representative fillings. The low-temperature behavior depends sensitively on the degree of commensuration. }}
    \label{fig:Res} 
\end{figure*}

Let us now turn our attention to lower temperatures ($T\lesssim V$), where for a range of fillings we also find evidence of $\rd\sim T$; see Fig.~\ref{fig:Res}(b) . However, there appear to be two distinct regimes with a $\rd\sim T$ behavior at high$-T$ (disordered) and low$-T$ (liquid-like), respectively, connected by a smooth crossover around $T\sim0.2V$. For densities $n\in(1/3,2/3)$, the crossover has a clear non-monotonic feature where the resistivity first has a small insulating-like upturn, followed by a drop, as a function of decreasing temperatures.\footnote{\textsf{This feature is insensitive to changes in system sizes as we go from  18$\times$18 sites to our largest system with 30$\times$30 sites.}} The upturn becomes more pronounced as the filling approaches $1/3$ and $2/3$, thereby suggesting its possible connection with the proximity to the insulating transitions. Conversely, our numerical results indicate a gradual disappearance of this upturn as the electron/hole filling becomes increasingly dilute, i.e. $0<n(1-n)<1/3$. These observations suggest an underlying connection between the resistivity upturn and the strength of correlations in the fluid phase. The approximate constraint of two frustrated bonds per elementary triangle leads to a reduced mobility of charges as only certain correlated hops are allowed. However, at dilute fillings, the constraint is lifted and the charge carriers become free to hop without a stark increase in the resistivity as a function of decreasing temperature.

The low$-T$ linear in temperature resistivity at half-filling is interesting for the following additional reasons. As a result of the underlying  particle-hole symmetry, there is a complete decoupling of charge and heat diffusion (see Appendix \ref{ap:TE}). Therefore, applying the Einstein-Nernst relation, $\sigma_{\tn{dc}}=D_c\chi$, with a compressibility that exhibits a diverging behavior $\chi \sim T^{-1}$ down to $T\sim O(V')$ leads us to conclude that $D_c$ is nearly temperature independent. Thus we have an interesting example of a ``low'' temperature $T$-linear resistivity that arises from the non-trivial behavior of the compressibility. The latter is a consequence of the geometric frustration of the particular model being studied here. Note that this is quite distinct from most other solvable examples of $T$-linear resistivity that arise at low-temperatures, where the compressibility is temperature independent while the diffusion coefficient (or equivalently, the scattering time) scales as $1/T$.

In the same low-temperature regime ($T\lesssim V$), we observe broadly two distinct types of behaviors. For our specific choice of $V'/V$ and for the range of temperatures that we have analyzed, the conductivity eventually vanishes near $n=1/3,~2/3$ below $T\sim 0.2V$ as the electrons crystallize into the insulating states, shown in Fig.~\ref{fig:phaseDiag}. For the incommensurate fillings on the other hand, we observe metallic transport down to the lowest temperatures. The resistivity $\rd$ scales roughly linearly with temperature. As described earlier, the system can crystallize into a host of other insulating states at a sequence of commensurate fillings $n=p/q$, depending on the ratio $V'/V$ and we expect the temperature dependence of $\rd$ to reflect these freezing transitions out of the liquid. The density dependence of the resistivity in the liquid-like regime is thus in marked contrast to the absence of any density dependence in the high-temperature (compressibility-dominated) regime. The strong density dependence in the low-temperature regime is naturally expected due to the incipient freezing transitions into the insulators at commensurate fillings.

\begin{figure}
\centering
\includegraphics[width=1\linewidth]{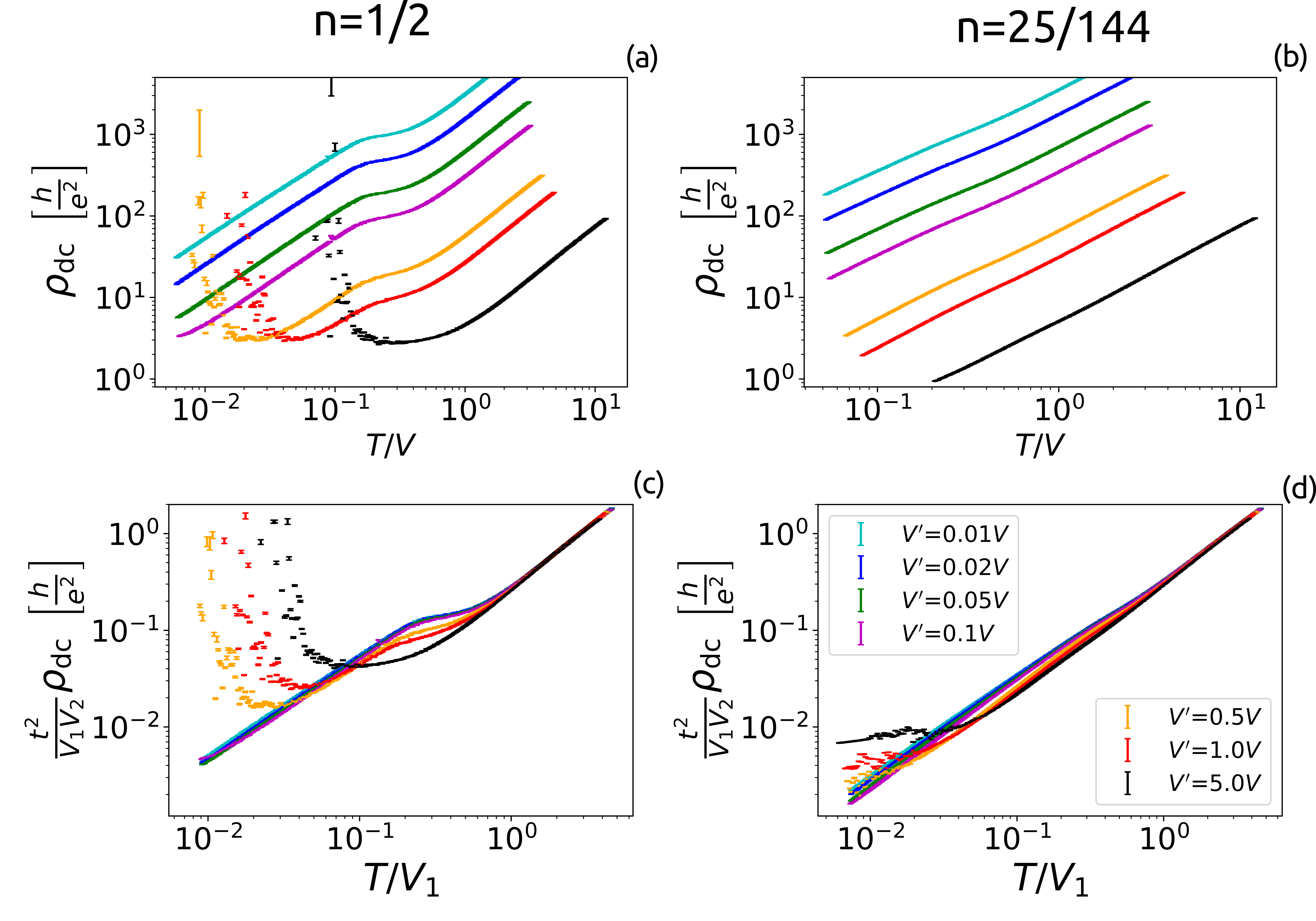}
\caption{\textsf{(a) $\rd$ as a function of temperature for varying $V'$ at half-filling (see inset of (d) for $V'-$legend). The transition temperature to the insulating stripe phase is controlled by $V'$. (b) $\rd$ as a function of temperature for varying $V'$ at $n=25/144$. There are no insulating transitions down to the lowest temperatures considered. Scaling collapse of the resistivity at (c) half-filling, and, (d) $n=25/144$.}}
\label{fig:ResVscaling} 
\end{figure} 

We can now make an interesting observation, based on purely dimensional grounds, on the form of the resistivity. Within our perturbative treatment of the nearest-neighbor electron hopping (and the associated current operator), it is clear that $\sigma\propto t^2$. Moreover, in two spatial dimensions, the conductivity has units of $e^2/h$. Thus, we can express the optical conductivity as,
\beq
\sigma(\omega)=\frac{e^2}{h}\frac{t^2}{V^2}~ {\cal{F}}\left(\frac{\omega}{V},\beta V,n \right),
\eeq
where ${\cal{F}}(...)$ is an arbitrary dimensionless function of the arguments listed as dots. Note that we have implicitly assumed $V'/V$ to be finite and fixed, when we rely on our non-degenerate perturbation theory. Comparing the above form with the explicit expression in Eqn.~\ref{eqn:optcond} can help us identify the behavior of the function ${\cal{F}}(...)$ as a function of its arguments. It now follows that the above form also fixes the slope of the resistivity in a regime displaying $\rd\sim T$ behavior,
\beq
\rd =\frac{h}{e^2} \widetilde{{\cal{F}}}(n)\frac{T}{T_0},
\label{eqn:Tlin}
\eeq
where $T_0=t^2/V$ and $\widetilde{{\cal{F}}}(n)$ is a dimensionless function of the filling. At a fixed (incommensurate) filling, the slope is determined by $T_0$ --- an emergent temperature scale in the problem that is much smaller than the relevant UV scales (i.e. $T_0\ll t,~V$). It is important to note that the expression above holds irrespective of whether $T\gg V$ or $T\ll V$, as long as $\rd\sim T$; the form of the function $\widetilde{{\cal{F}}}(n)$ on the other hand can certainly be different. Since $T\gg T_0$ by construction in our approach, we naturally expect to obtain bad-metallic transport. It is important to note that the parametric dependence of the resistivity here is markedly different from the weak-coupling limit describable within Landau-Boltzmann theory. 

Let us now turn to our results for the resistivity and verify the extent to which the above scaling form describes the data. Recall that the finite resistivity (and scattering rate) in our description of transport arises due to a $V'>0$. For commensurate fillings, the linear in $T$ behaviour of $\rd $ is expected to be cutoff by a transition to insulating states at $T\sim O(V')$. For the case of $n=1/2$, this is illustrated in Fig.~\ref{fig:ResVscaling}(a). We also show the results for $n=25/144$ in Fig.~\ref{fig:ResVscaling}(b), where such freezing transitions are avoided for the specific values of $V'$. To illustrate the scaling collapse, we find it convenient to express the result in terms of the nearest and next nearest-neighbour interactions $V_1$ and $V_2$, respectively. The scaling forms are similar to Eqn.~\ref{eqn:Tlin}, but with $T_0\sim t^2/V_1 V_2$. The scaling collapse for $\rd$ for the two fillings considered above are shown in Fig.~\ref{fig:ResVscaling} (c), (d), respectively. The high-temperature regime shows nice scaling collapse irrespective of the relative ratio of $V'/V$. For  commensurate fillings (e.g. $n=1/2$), as long as the temperature of the freezing transition ($\sim O(V')$) is much smaller than $V$, a wide intermediate regime of $T-$linear resistivity can arise that exhibits scaling collapse of the form discussed above. We discuss the thermoelectric coefficients for our model in Appendix \ref{ap:TE}.

We end this section by noting that the $T-$linear resistivity discussed in Eqn.~\ref{eqn:Tlin} is superficially similar in form to the results obtained for the resistivity for lattice SYK models, as summarized in the introduction. However, the underlying mechanism leading to $\rd\sim T$ is strikingly distinct between the two approaches. In our triangular-lattice models, the temperature dependence of $\rd$ appears to arise primarily from the compressibility (and not the scattering rate), while for the SYK models, the compressibility is temperature independent and the scattering rate scales as $1/T$.

\section{\textsf{Outlook}}
\label{sec:outlook}

Describing the transport properties of strongly interacting quantum matter at a finite temperature and for generic densities is in general a difficult problem. In this work, we have identified and solved the problem of coupled charge and heat transport for a family of strongly interacting models with geometric frustration on the triangular lattice. Using a combination of numerical Monte Carlo simulations, analytical linear response calculations for transport, and analytical insight in an interacting, but solvable, limit, we have identified a number of interesting results for bad metallic behavior over a broad range of temperatures. In particular, in the limit of strong interactions we find resistivities that scale linearly with temperature over a broad intermediate range of temperatures and fillings, with magnitude much larger than $h/e^2$; these results are qualitatively distinct from the behavior of a quasiparticle-based Boltzmann equation approach for the same model at weak interactions.  

One of the motivations for studying the models introduced in this paper arises from their direct experimental relevance for describing a class of new moir\'e transition metal dichalcogenide (TMD) van der Waals materials. As a result of the natural lattice mismatch between distinct TMD materials, their heterobilayers (e.g. WSe$_2$/WS$_2$) form large-period triangular moir\'e superlattices even at zero-twist angle. The resulting moir\'e bands are well separated from the other dispersive bands and the carrier density can be tuned easily using an external gate voltage. This platform is ideal for simulating single-band Hubbard-model physics on the triangular lattice \cite{McDonald2018}, where the {\it valley} degree of freedom plays the role of an effective spin. However, in addition to an on-site Coulomb repulsion, the long-range Coulomb interactions play a defining role in this system. Recent experiments have estimated the ratio of on-site interactions to the nearest neighbor hopping to be as large as $20$ in a certain window of external parameters \cite{JieTMD2020,FengwangTMD2019}.
In such a limit, the electronic states at a filling factor of $n\leq 1$ are effectively described by spinless fermions with an occupancy of either $0$ or $1$ at every site and the ground state properties are determined by the long-range Coulomb interactions. {\footnote{\textsf{We note that the spin degree of freedom can not be ignored completely here and does enter in the description of the true ground state.}}} Remarkably, a plethora of Wigner crystalline states at commensurate fillings, $n=p/q$, have been observed experimentally in this heterobilayer system \cite{xu2020abundance}, many of which can be described by the energetics of the purely classical problem with Coulomb interactions using MC simulations. In the limit of low temperatures, the experimental ground state as a function of filling \cite{xu2020abundance} resembles the schematic phase-diagram in Fig.~\ref{fig:fig1}b, where the precise orderings at commensurate densities are determined by the longer ranged interactions and the ``correlated fluid'' corresponds to a metallic regime. At temperatures that are comparable to or larger than the bandwidth of the flat-bands but significantly smaller than the gap to the higher energy bands, the computations discussed in the present paper can capture some aspects of the transport behavior as a function of decreasing temperature due to the leading quantum mechanical corrections from the electron hopping processes in the limit of strong Coulomb interactions. Detailed measurements of the electronic compressibility and dc transport as a function of temperature and filling in future experiments on TMD heterobilayers are thus highly desirable.   

Measurements of the charge diffusivity and compressibility in a cold-atoms based
realization of the on-site Hubbard model on the two-dimensional square lattice with interaction comparable to the bandwidth \cite{Bakr} found evidence of bad-metallic behavior and a broad regime of $T-$linear resistivity from temperatures higher than the bandwidth down to lower temperatures. Interestingly, while the compressibility and diffusivity exhibit complicated temperature dependent crossovers, the resistivity as determined using the Einstein relation does not exhibit any signs of a crossover as a function of decreasing temperatures. The predictions for the model studied in this paper in a distinct microscopic setting are markedly different, in terms of the origin of $T-$linear resistivity as well as the various temperature dependent crossovers, both of which are intimately tied to the geometric frustration in the Ising-limit. In spite of the possible experimental challenges associated with engineering the triangular lattice  models studied here (even in the absence of the longer-ranged interactions) using cold-atoms, it will be interesting to probe the compressibility and diffusivity directly in future experiments, especially in the regime where the hopping strength is comparable to interactions. Given that the statistics of the excitations that contribute to transport is likely irrelevant for the intermediate scale transport properties in the regime discussed in this paper, measuring the spin-diffusion coefficient \cite{Zwierlein} for the XXZ spin model on the triangular lattice over a broad intermediate range of temperatures in future experiments is equally desirable. 

Finally, we note that the model defined in terms of hard-core bosons at half-filling does not suffer from the infamous sign-problem and can therefore be solved exactly using quantum Monte Carlo simulations. It will be interesting to probe signatures of a possibly bad-metallic regime in the limit of strong interactions over an extended range of temperatures using various ``proxies'' for resistivity \cite{Lederer2017} in the absence of numerical data for real frequencies. Moreover, the quantum Monte Carlo calculations can shed new light on the model at intermediate coupling and on the coherence scale associated with a crossover out of the bad metallic regime; these are questions that lie beyond the scope of our present perturbative approach. On the other hand, for the fermionic model with only nearest neighbor interactions, a two-site dynamical mean-field theory \cite{DMFT} computation can offer complementary insight without having to resort to the specific perturbative limit considered in this paper. 

\section*{\textsf{Acknowledgements}}
We thank E. Berg, V. Elser, J. Hofmann, C. Mousatov, E. Mueller, T. Senthil and J. Sethna for useful discussions and S. Musser for comments on an earlier version of this manuscript. J.F.M.V. is supported by a Cornell graduate fellowship. DC is supported by faculty startup funds at Cornell University.

\begin{appendix}

\section{\textsf{Monte-Carlo simulations}}
\label{ap:MC}
In this appendix, we provide additional details of our Monte-Carlo computations. In order to simulate the system described by Eqns.~\ref{eqn:hamiltonian} and \ref{eqn:LongV} in the main text, we used the standard Metropolis algorithm along with the Glauber acceptance rule \cite{Glauber}. For our classical model, every MC move consists of selecting a site at random and modifying (preserving) its occupation number with probability $P_\r$ ($\overline{P}_\r$), where $P_\r/\overline{P}_\r = e^{\beta \left( \varepsilon_\r -\mu \right) \delta n_\r  }$, subject to $P_\r+\overline{P}_\r = 1$. Here, $\delta n_\r=\text{mod} \left(n_\r +1,2 \right) -n_\r$, is the attempted change in the occupation at site $\r$ and $\varepsilon_\r $ is the local energy defined in Eqn.~\ref{eqn:loc_ene}. The above rule guarantees that at high-temperatures ($T\gg V$) there is an equal probability of occupying or vacating a site. Then, for an $L\times L$ system, each step of the algorithm  consists of $L^2$ MC moves.  Throughout this procedure, we impose periodic boundary conditions and use the ``minimum-image'' convention on the hexagonal unit cell to calculate distances between all pairs of points \cite{SimulationLiq}.

In a given run of the simulation, the system starts from a high temperature in the range $T\sim 10-50~V$, which is then progressively lowered. We use the Monte Carlo algorithm with the Glauber rule to sample the Gibbs distribution and calculate all thermodynamic and transport observables at a given temperature. To avoid spurious correlation induced freezing in our simulations, we gradually decrease the temperature in steps of a small $\Delta T/T\ll 1$. This cooling rate allows for sweeping across a broad temperature range while still having a high-resolution grid at low temperatures. At each fixed temperature, the computation proceeds as follows: (i) We use a parent configuration for the system and  use a binary search algorithm to obtain the chemical potential that lets us achieve the desired density at that temperature. (ii) With the fixed chemical potential at this temperature, the MC sampling proceeds by performing $1000$ thermalization steps, followed by $25000$ sampling steps, where each step of the algorithm  consists of $L^2$ MC moves. The number of thermalization steps was determined empirically by analyzing the equilibration timescale associated with the energy and particle number.

\begin{figure}
\centering
\includegraphics[width=1\linewidth]{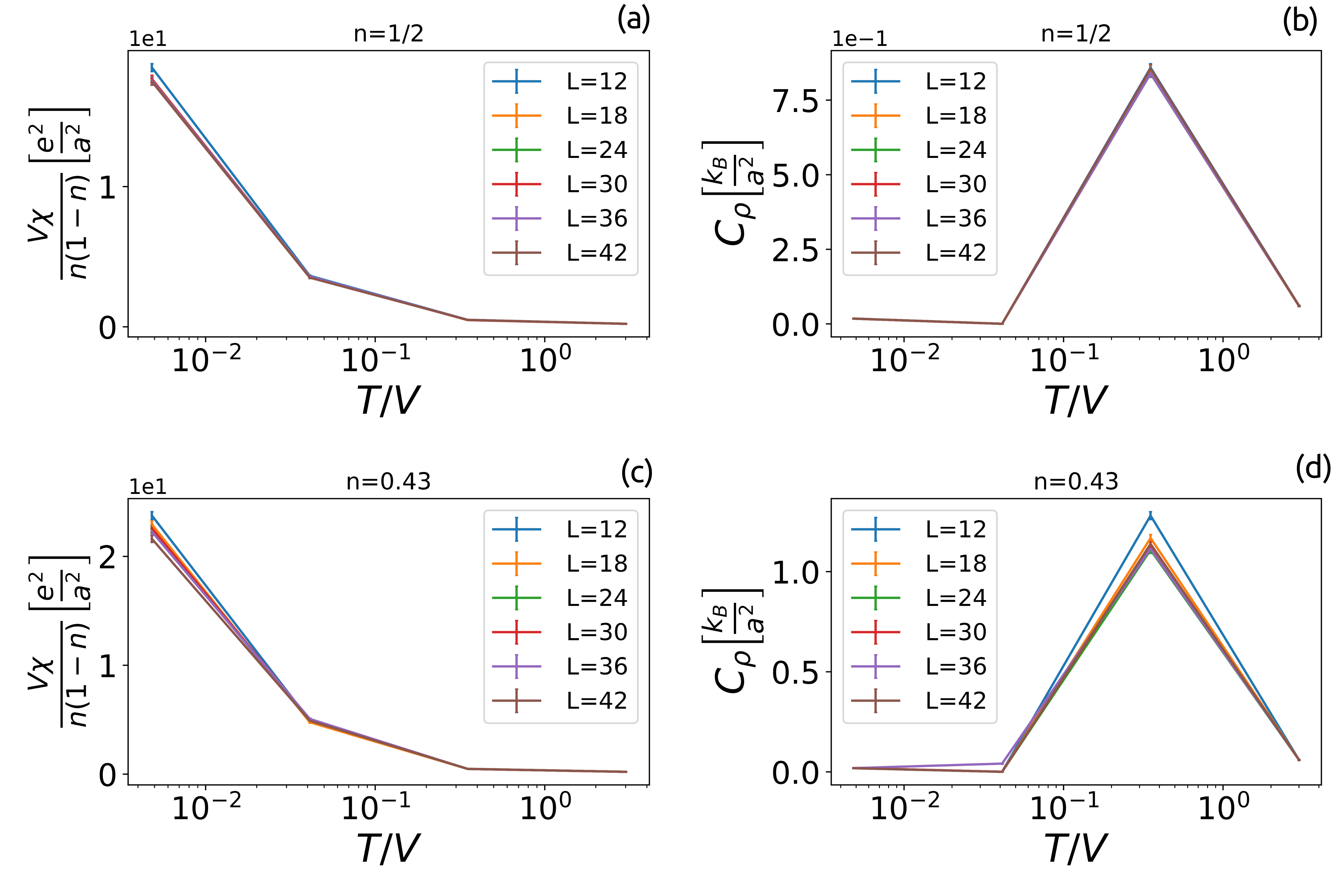}
\caption{\textsf{Effects of finite size effects for different thermodynamic quantities as a function of filling and temperature.}}
\label{fig:finite-size} 
\end{figure} 

In each of the sampling steps, we store the particle number and the total energy of the configuration. Additionally, we record the quantities needed for evaluating the currents and the local energies, as discussed in the main text. 

Once the MC sampling is completed, we calculate the different thermodynamic quantities using the standard relations:
\beq
\chi=\frac{e^2 \beta}{ \tn{vol} }\left(\left\langle N_{\{n\}}^{2}\right\rangle -\left\langle N_{\{n\}}\right\rangle ^{2}\right),
\label{eq:flucreschi2}
\eeq
\beq
c_\mu=\frac{k_B \beta^2}{ \tn{vol} }\left(\left\langle E_{\{n\}}^{2}\right\rangle -\left\langle E_{\{n\}}\right\rangle ^{2}\right),
\label{eq:flucresc}
\eeq
\beq
\zeta=\frac{k_B e \beta^2}{ \tn{vol} }\left(\left\langle N_{\{n\}}E_{\{n\}}\right\rangle -\left\langle N_{\{n\}}\right\rangle \left\langle E_{\{n\}}\right\rangle\right),
\label{eq:flucresz}
\eeq
and 
\beq
c_{\rho}=c_{\mu}-T\frac{\zeta^{2}}{\chi}.
\eeq
In order to determine the uncertainties associated with these quantities, we perform bootstrap resampling \cite{degroot2012probability}.

Finally, we also studied the finite-size effects in our simulations. We compared the compressibility, $\chi$, and the heat capacity at fixed density, $c_\rho$, for different system sizes as a function of temperature; the results are shown in Fig.~\ref{fig:finite-size}. We note that the results appear to have converged already for system sizes of $30\times30$, which is used throughout this work. 

\section{\textsf{Linear response theory for transport}}
\label{ap:LR}
The particle and energy currents in the system can be obtained from the respective continuity equations. 
This leads to the particle current,
\beq
 \bm{J} & =-\frac{it}{\hbar}\sum_{\langle \r\r'\rangle }\bm{D}_{\r\r'}c_{\r}^{\dagger}c_{\r'},
\eeq
where $\bm{D}_{\r\r'}=\r-\r'$. (The charge current is then $e\bm{J}$.) Similarly, the energy current is given by (after ignoring a term proportional to $O(t^2)$),

\beq
\bm{J}_{E} 
 &= -\frac{i}{2\hbar}\sum_{\r\r'}\left[\bm{D}_{\r\r'}\left(\varepsilon_{\r'}+\varepsilon_{\r}\right)+\bm{\epsilon}_{\r}-\bm{\epsilon}_{\r'}\right]t_{\r\r'}c_{\r}^{\dagger}c_{\r'},
\eeq
where $\varepsilon_{\r}$ was defined in Eqn.~\ref{eqn:loc_ene} above and 
\beq
\bm{\epsilon}_{\r}=\sum_{\r''}\bm{D}_{\r''\r}V_{\r\r''}n_{\r''},
\label{eqn:enecurr}
\eeq
with $V_{\r,\r'}=V\delta_{|\r-\r'|,a}+V^{\prime}\frac{e^{-\left|\r-\r'\right|/\ell}}{\left|\r-\r'\right|}$. The term in Eqn.~\ref{eqn:enecurr} describes the part of the energy current which arises only in the presence of extended interactions and is necessary for recovering the continuity equation for the energy current.  
The heat current to leading order in $t$ is thus defined as, 
\beq
\bm{J}_{Q} &=& \bm{J}_{E}-\mu\bm{J}\\
 &=& -\frac{it}{2\hbar}\sum_{\left\langle \r,\r'\right\rangle }\left[\bm{D}_{\r\r'}\left(\varepsilon_{\r'}+\varepsilon_{\r}-2\mu\right)+\bm{\epsilon}_{\r}-\bm{\epsilon}_{\r'}\right]c_{\r}^{\dagger}c_{\r'}.
\eeq

We are interested in coupled heat and charge transport within linear response theory \cite{ziman}. The charge conductivity was introduced explicitly in Eqn.~\ref{cond} in the main text. We can analogously define the following (longitudinal) thermoelectric conductivities:
\beq
\alpha^{xx} &\equiv& \frac{\beta}{\tn{vol}} \int_0^\infty d\tau~e^{i\omega^+\tau} \int_0^\beta d\lambda \langle J^x(\tau-i\lambda) J_Q^x\rangle,\\
\bar{\kappa}^{xx} &\equiv& \frac{\beta}{\tn{vol}} \int_0^\infty d\tau~e^{i\omega^+\tau} \int_0^\beta d\lambda \langle J_Q^x(\tau-i\lambda) J_Q^x\rangle,
\label{thermoel}
\eeq
where the expectation values are evaluated as earlier, $\langle...\rangle = \tn{Tr}(...e^{-\beta H_{\tn{eff}}}/Z)$.

In terms of these quantities, the thermopower and the Lorentz ratio are defined in the usual way,
\beq
S &=& \frac{\alpha}{\sigma},\\
L &=& \frac{\kappa}{T\sigma},
\eeq
where the open-circuit thermal conductivity, $\kappa$, is related to $\bar{\kappa}$ as,
\beq
\kappa=\bar{\kappa}-T\frac{\alpha^{2}}{\sigma}.
\eeq

These two-point correlations of the different current operators (related to the Onsager coefficients) can be recast in the following form using the spectral representation,
\beq
L_{ab} &=& \frac{\pi\left(1-e^{-\beta\hbar\omega}\right)}{\omega}\frac{1}{\tn{vol}}\sum_{\left\{ n,n^{\prime}\right\} }\frac{\exp\left(-\beta E_{\{n\}}\right)}{Z} |J_{a,b}|^2_{n,n'} \delta\left(\hbar\omega+E_{\left\{ n\right\} }-E_{\left\{ n^{\prime}\right\} }\right), \label{spec}\\
|J_{a,b}|^2_{n,n'} &\equiv& \langle\left\{ n\right\} |J_a|\left\{ n^{\prime}\right\} \rangle\langle\left\{ n^{\prime}\right\} |J_b|\left\{ n\right\}\rangle.
\eeq
In the above expression, $a,b=\left\{ 1,2\right\}$ with $J_{1}=J,~J_{2}=J_{Q}$; $|\left\{ n\right\} \rangle$ represents the set of configurations in the basis of occupation numbers associated with the effective Hamiltonian (for $t=0$) and $E_{\{n\}}$
are the corresponding energies. For the sake of notational convenience, we have dropped the spatial indices associated with the longitudinal response above. It is useful to express the charge and energy currents using the following compact notation:
\beq
\bm{J}_{a}=-i\sum_{\left\langle \r,\r'\right\rangle }\bm{Y}_{\r\r'}^{a}c_{\r}^{\dagger}c_{\r'},
\eeq
where $\bm{Y}_{\r\r'}^{a}$ is antisymmetric in $\r,\r'$ and can be read off for the charge and heat current from the expressions introduced earlier. The first step associated with the computation in Eqn.~\ref{spec} involves computing the matrix elements of the current operators between different configurations. Clearly, these matrix elements are finite only when the single-electron hops between nearest neighbor sites relate the different configurations in the basis of occupation numbers.

After simplifying these matrix elements, we arrive at the following expression:
\beq
L_{ab} 
  &=&\frac{\pi \left(1-e^{-\beta\hbar\omega}\right)}{\omega}\frac{1}{\textnormal{vol}}\left\langle \sum_{\left\langle \r\r'\right\rangle } w^{ab}_{\r\r'}~ n_{\r}\left(1-n_{\r'}\right)\delta\left(\hbar\omega+\Delta \ve_{\r\r'}\right)\right\rangle,\\
  w^{ab}_{\r\r'}&=& Y_{\r\r'}^a Y_{\r\r'}^b,\\
  \Delta \ve_{\r\r'}&=& E_{\{n_{\r_{1}}\cdots,n_{\r},\cdots,n_{\r'},\cdots, n_{\r_{N}}\}}-E_{\{n_{\r_{1}},\cdots,\overline{n_{\r}},\cdots,\overline{n_{\r'}},\cdots,n_{\r_{N}}\}},
\eeq
where for the longitudinal response, we have assumed $Y = [\bm{Y}]^x$. 

Recall once again that all thermal expectation values are evaluated in the purely classical theory with $t=0$. $\Delta \ve_{\r\r'}$ corresponds to the difference in the energy of the system when an electron hops from an occupied site $\r$ to an unoccupied site $\r'$ (i.e. subject to the constraint, $n_{\r}\left(1-n_{\r'}\right)$) with $\overline{n}_{\r}\equiv 1-n_{\r}$. In practice, the sum over the nearest neighbours in the equation above is carried out for a single configuration of the system in the MC calculation, $\{n \}$. Therefore, it is useful to recast the energy difference making reference only to the initial state before the electron performs the hop. We now recall the definition for $E_{\{n\}}$ in terms of the local energies,  $\varepsilon_{\r}^{\{n\}}$, as introduced in Eqn.~\ref{En}. The energy difference can then be simplified to:

\beq
\Delta\varepsilon_{\r\r'}=\left(\overline{n}_{\r}-n_{\r}\right)\varepsilon_{\r}^{\left\{ n\right\} }+\left(\overline{n}_{\r'}-n_{\r'}\right)\varepsilon_{\r'}^{\left\{ n\right\} }+\left(\overline{n}_{\r}-n_{\r}\right)\left(\overline{n}_{\r'}-n_{\r'}\right)V_{\r,\r'}.
\eeq

Moreover, by imposing the additional constraint associated with the allowed electron hops with $n_{\r}\neq n_{\r'}$, implying $\overline{n}_{\r}=n_{\r'}$ and $\overline{n}_{\r'}=n_{\r}$, we have

\beq
\Delta\varepsilon_{\r\r'}
 =\left(n_{\r'}-n_{\r}\right)\left(\varepsilon_{\r}^{\left\{ n\right\} }-\varepsilon_{\r'}^{\left\{ n\right\} }\right)-\left(n_{\r'}-n_{\r}\right)^{2}V_{\r,\r'}.
\eeq
This readily leads to the expression for the optical conductivity in Eqn.~\ref{eqn:optcond} and Eqn.~\ref{eqn:spec_weight} upon identifying $Y_{\r\r'}$ with the charge current.

Similarly, we can now obtain the other conductivities defined in Eqn.~\ref{thermoel} in a straightforward fashion by identifying the $Y_{\r\r'}^{a,b}$ with either charge or heat currents, respectively. Thus,

\begin{align}
\alpha^{xx}\left(\omega\right) & =\beta  \frac{e}{h}t^2\frac{\left(1-e^{-\beta\hbar\omega}\right)}{\hbar\omega}\sum_{\{n\}}\frac{e^{-\beta H_{\tn{eff}}}}{Z}  \nonumber \\
 & \times \frac{2\pi^{2}}{\textnormal{vol}}\sum_{\left\langle \r\r'\right\rangle }\left[\bm{D}_{\r\r'}\right]^{x}\left[\bm{D}_{\r\r'}\left(\varepsilon_{\r'}+\varepsilon_{\r}-2\mu\right)+\bm{\epsilon}_{\r}-\bm{\epsilon}_{\r'}\right]^{x}\Delta{}_{\r\r'}\left(\omega\right),
\end{align}

\begin{align}
\bar{\kappa}^{xx}\left(\omega\right) & =  \frac{\beta}{h}t^2\frac{\left(1-e^{-\beta\hbar\omega}\right)}{\hbar\omega}\sum_{\{n\}}\frac{e^{-\beta H_{\tn{eff}}}}{Z}  \nonumber\\
 & \times \frac{2\pi^{2}}{\textnormal{vol}}\sum_{\left\langle \r\r'\right\rangle }\left(\left[\bm{D}_{\r\r'}\left(\varepsilon_{\r'}+\varepsilon_{\r}-2\mu\right)+\bm{\epsilon}_{\r}-\bm{\epsilon}_{\r'}\right]^{x}\right)^{2}\Delta{}_{\r\r'}\left(\omega\right).
\end{align}
The numerical results for these thermoelectric conductivities are computed in the dc limit, $\omega\rightarrow0$, by adopting the same binning procedure we adopted for the charge conductivity.

In particular, the dc limit corresponds to the value of the height of the averaged histograms for the MC configurations at the smallest sampled frequency. 

\section{\textsf{Thermoelectric coefficients}}
\label{ap:TE}

In the main text, we focused exclusively on charge transport via electron hops in the small $t$ regime. However, it is important to study the effects of both charge and heat transport, which are coupled together via the thermoelectric coefficients in this strong-coupling regime. The coupled diffusion equations can be expressed as, 
\beq
\frac{\partial^2 \mathbf{n}}{\partial t^2} = \mathbf{D} \nabla^2 \mathbf{n},
\eeq
where $\mathbf{n}=(n, \varepsilon -\mu n)$, with $\varepsilon$ the local energy density and $ \mathbf{D}$, the diffusion matrix. 

\captionsetup[figure]{justification=centerlast}
 \begin{figure*}[t!]
    \centering
        \begin{subfigure}{0.49\textwidth}   
            \centering 
            \includegraphics[width=\textwidth]{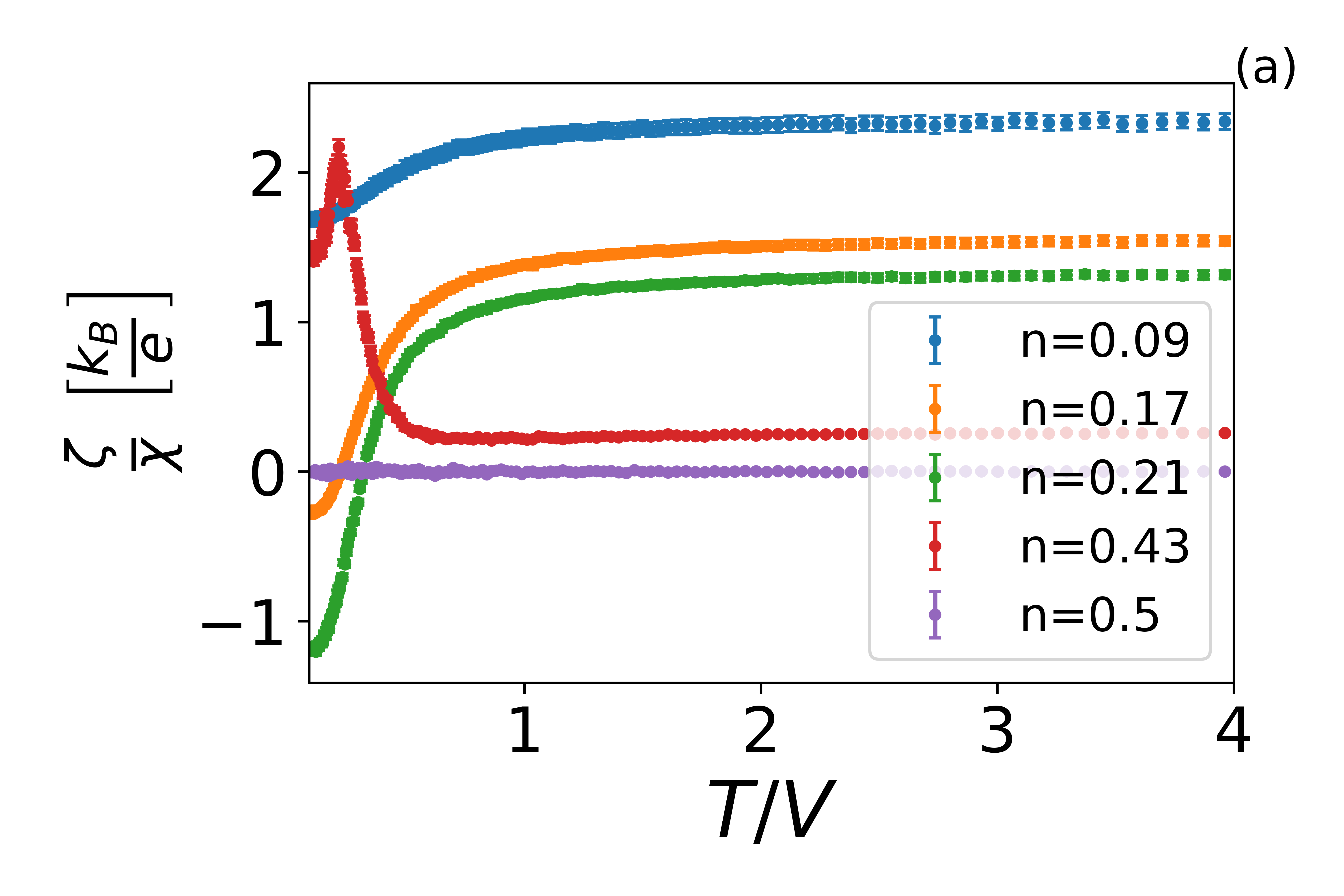}
            
        \end{subfigure}
        ~
        \begin{subfigure}{0.49\textwidth}   
            \centering 
            \includegraphics[width=\textwidth]{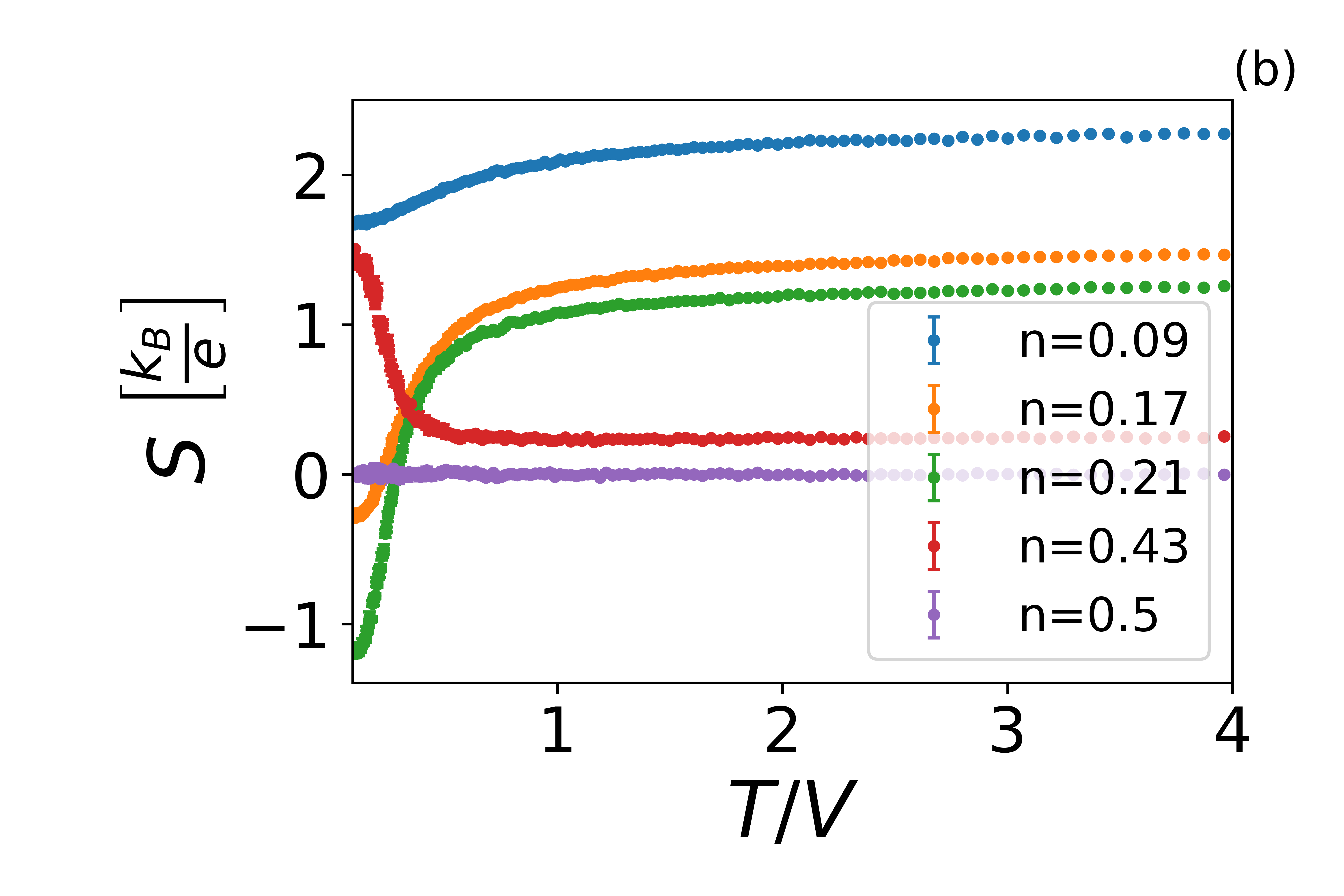}
        \end{subfigure}
         \vskip\baselineskip
          \begin{subfigure}{0.49\textwidth}   
            \centering 
            \includegraphics[width=\textwidth]{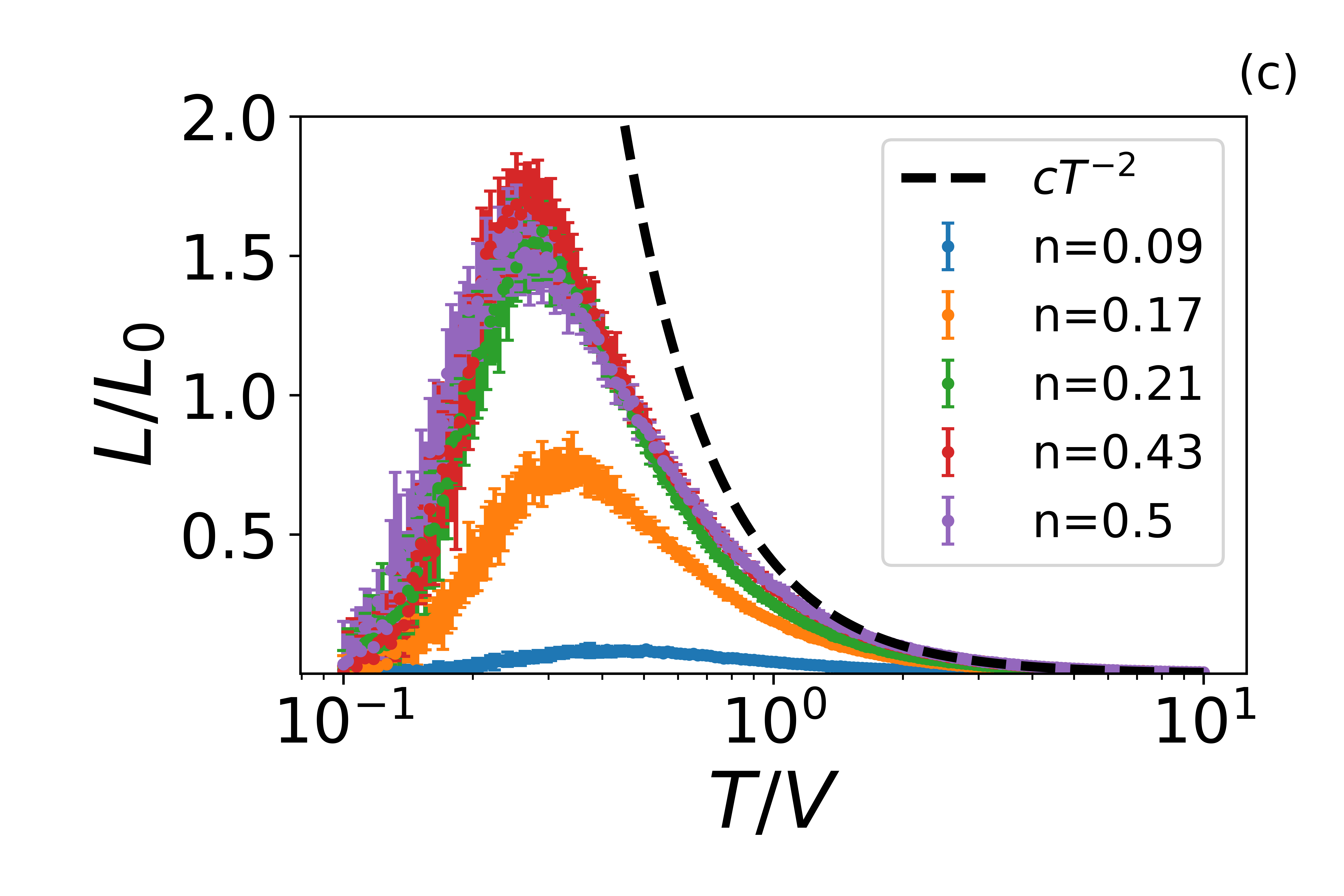}
        \end{subfigure}
     \caption{\textsf{Thermodynamic and transport coefficients that contribute to thermoelectric coupling. See text for definition of individual quantities.}}%
    \label{fig:thermoE}%
\end{figure*}

The eigenvalues, $D_\pm$, of the diffusion matrix satisfy,
\begin{align}
D_+ D_- &= D_c D_Q, \label{eqn:dete}\\
D_+ + D_- &=D_c + D_Q \left( 1+ \frac{(S-\zeta  / \chi )^2}{L} \right), \label{eqn:trace}
\end{align}
where $\chi,\zeta$ are as defined earlier and $D_c=\sigma/\chi$ ($D_Q=\kappa /c_{\rho}$) is the charge (heat) diffusion constant in the absence of any coupling between the two. The thermal conductivity, $\kappa$, is defined with open boundary conditions (as introduced earlier), $L=\kappa/T\sigma$ is the Lorenz number, and $S=\alpha/\sigma$ is the thermopower. 

At $n=1/2$, both $S=\zeta=0$ (see Fig.~\ref{fig:thermoE}a, b),  such that we recover $D_{-}=D_c$ and $D_{+}=D_Q$. 

We can determine the strength of the thermoelectric coupling by studying the coefficient that multiplies $D_Q$ in the trace of the diffusion matrix in Eqn.~\ref{eqn:trace}: 
\beq{}
\delta D_{Qc} \equiv \frac{(S-\zeta  / \chi )^2}{L} .
\label{eqn:dQC}
\eeq{}
To characterize the behavior of $\delta D_{Qc}$ as a function of filling and temperature, we use classical MC to calculate all of the individual contributions; the results are shown in Fig.~\ref{fig:thermoE}. Away from half-filling and in the limit of both high and low temperatures, $S$ and $\zeta/\chi$ approach a constant value asymptotically. 

For $T\gg V$, the asymptotic behavior is well described by Heikes formula \cite{ChaikinThermopower} and nominally suggests a (partial) decoupling of charge and heat diffusion. However, the behavior of the Lorentz ratio changes this expectation drastically. We observe a strong deviation away from the Widemann-Franz law, which is expected since heat and charge conductivity in this regime does not arise from elastic scattering. Furthermore, we note that $ L/L_0 \sim 1/T^2$, were $L_0 = \frac{\pi^2}{3} \frac{k_B^2}{e^2} $. Similarly, we also find that the leading non-vanishing correction to $S-\zeta/\chi$ falls off as $1/T$. Therefore, both the numerator and denominator of $\delta D_{Qc}$ in Eqn.~\ref{eqn:dQC} go to zero in this high-temperature limit in the same fashion and the final result is a constant that only depends on the density. We are thus led to the conclusion that while the thermoelectric effects do not dramatically alter the temperature dependence of charge and heat diffusion constants, their density dependence can nevertheless be modified at high temperatures.

\captionsetup[figure]{justification=centerlast}
 \begin{figure*}[t!]
    \centering
        \begin{subfigure}{0.49\textwidth}
            \centering
            \includegraphics[width=\textwidth]{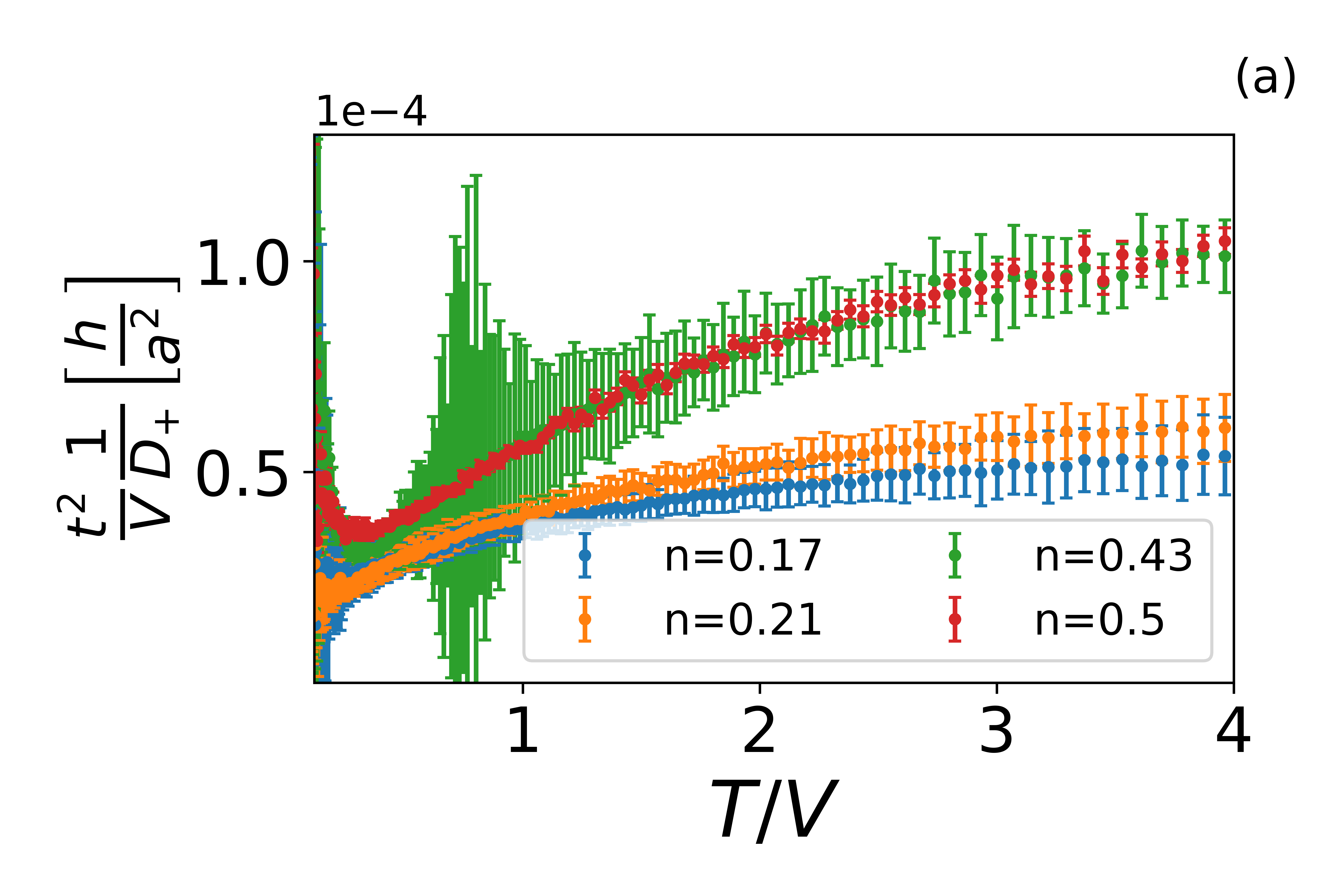}
            
        \end{subfigure}
        ~
        \begin{subfigure}{0.49\textwidth}  
            \centering 
            \includegraphics[width=\textwidth]{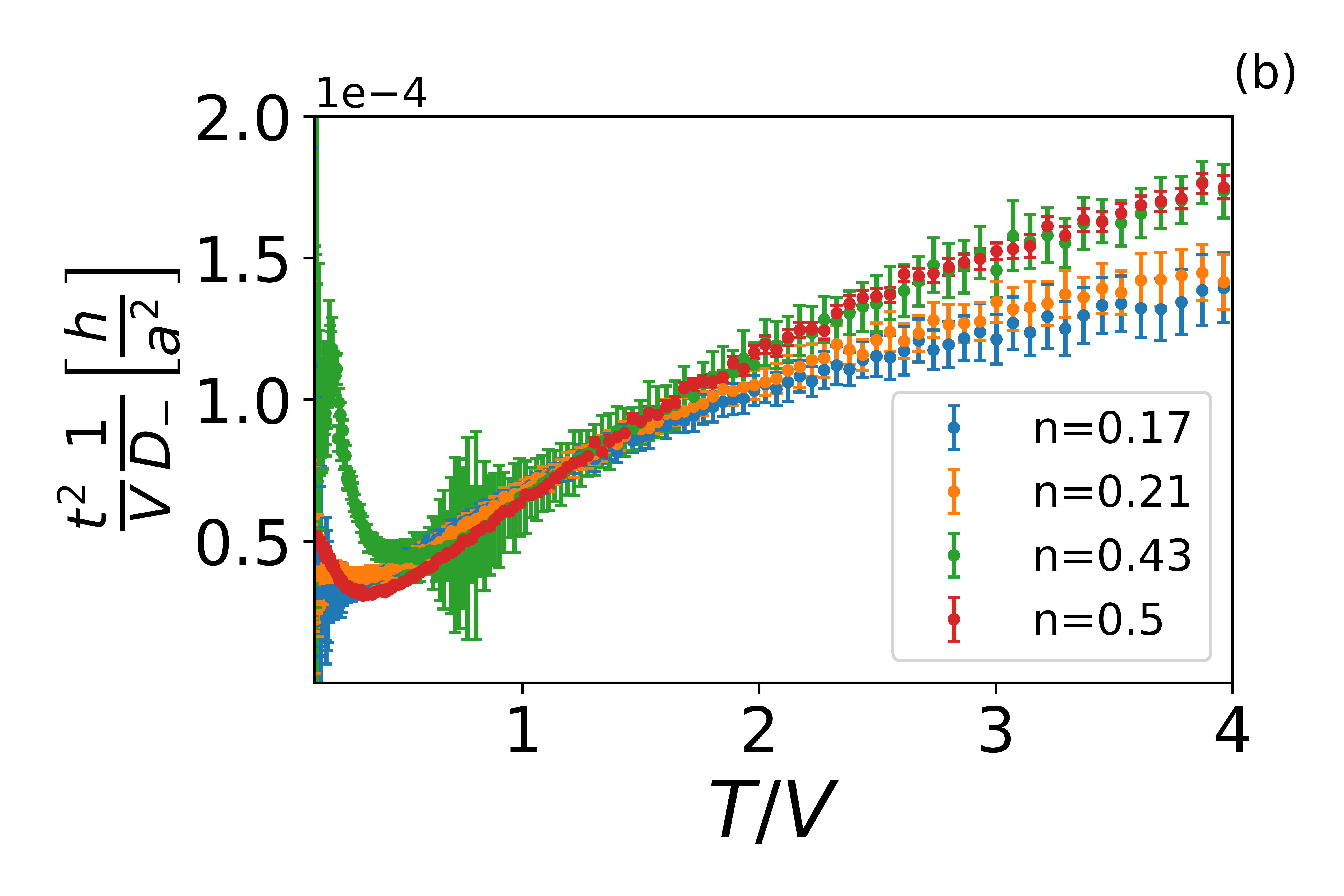}
          
        \end{subfigure}
       
     \caption{\textsf{Inverse eigenvalues of the diffusion matrix. Numerical errors prevent the accurate determination of these diffusion constants for dilute fillings and for low temperatures deep in the correlated regime. }}%
    \label{fig:diff-rate}%
\end{figure*}

The results for $D_{\pm}$ are displayed in Fig.~\ref{fig:diff-rate} for a wide range of temperature and fillings. As expected, both quantities display a saturation in the limit of high and low temperatures. As was already noted above, for $1/3 \lesssim n \lesssim 2/3 $ and $T\gg V$, the thermoelectric coupling in Eqn.~\ref{eqn:dQC} is small $\delta D_{Qc}\ll 1$. As a result, the charge and heat diffusion coefficients approximately decouple for these intermediate fillings in the high temperature limit. Upon moving further away from half-filling, the thermoelectric effects become important. In particular, at $T > V$ and $n\lesssim 1/3~ (\gtrsim 2/3)$, $\delta D_{Qc}$ grows with temperature until it saturates to a density-dependent value. At saturation, $D_{Qc}$ grows monotonically as we go away from half filling. Surprisingly, even for very dilute fillings $n\sim 1/6$, the contribution from the thermoelectric coupling is still small $\delta D_{Qc} \sim 0.2$. Consequently, $\mathbf{D}$ becomes approximately diagonal with $D_+ \sim D_Q$ and $D_- \sim D_c$. The use of the Nerst-Einstein relation is justified for a wide range of fillings provided $D_c \rightarrow D_-$ at $T\gg V$. However, $D_{\pm}$ is still sensitive to thermoelectric coupling at very dilute fillings $n\lesssim 0.1 (\gtrsim 0.9)$. For intermediate fillings, $D_-$ has the same density dependence as, $D_{c}\propto 1/n(1-n)$.

As the system approaches the correlated fluid regime at low temperatures, $D_c$ increases faster than $D_Q$ such that $D_+ \sim D_c$. At the point where the diffusion eigenvalues cross and become nearly degenerate, there is an increase in roundoff errors arising from the underlying cancellations, which become noticeable around $T\sim 0.8V$ for both $D_{+}$ and $D_{-}$ in Fig.~\ref{fig:diff-rate}. Further decrease in temperature leads to the onset of the constraint on the number of frustrated bonds on each fundamental triangle for $n\in (1/3,2/3)$. In this regime at $n=1/2$, the relative size of charge and heat diffusion follows from the fact that energy can only diffuse as charges perform correlated hops that maintain two frustrated bonds per triangle. However, particle hole symmetry forces $\delta D_{Qc}=0$. Overall, this implies that density gradients relax roughly on the same scale as energy gradients.

At the lowest temperatures in our simulation, the diffusion eigenvalues seem to approach a constant for all  densities. Here $D_{\pm}$ display large errors due to cancellations in our computation of $c_\rho$ and $\kappa$. These cancellations are a consequence of the increased correlation between charge and local energy fluctuations together with the suppression of total energy fluctuations.  
Despite these complications, a saturation of the diffusion coefficients for both $D_{\pm}$ hints strongly at a purely thermodynamic origin for the temperature dependence of transport, that is nevertheless fundamentally different from the statistical regime at high temperatures.

\section{Comparison with exact solution in the Ising limit}
\label{ap:Is}
In order to benchmark our MC algorithm, we compared our results with the exact solution obtained by Wannier. The exact solution at zero magnetic field (for Ising spins) was compared with the internal energy and the heat capacity obtained within our MC calculation at $\mu=0$. Note that the temperature has to be rescaled in appropriate units; the Ising interaction term, $J$ in Wannier's notation corresponds to the nearest-neighbor repulsion, $V/2$, in our notation. The internal energy of the system, $U$, is given by 
\beq
\frac{U}{N}=\frac{V}{1-\nu}\bigg[1-f_{2}\left[\nu\left(T\right)\right]~K\left[f_{3}\left(\nu\left(T\right)\right)\right]\bigg],
\eeq
where 
\beq
f_{2}\left(\nu\right) &=& \frac{8}{\pi}\frac{\nu\left(3-\nu\right)}{4\sqrt{\nu}+\sqrt{\left(\nu+1\right)^{3}\left(3-\nu\right)}},\\
f_{3}\left(\nu\right) &=& \frac{4\sqrt{\nu}-\sqrt{\left(\nu+1\right)^{3}\left(3-\nu\right)}}{4\sqrt{\nu}+\sqrt{\left(\nu+1\right)^{3}\left(3-\nu\right)}},\\
\nu\left(T\right) &=& 1+\text{tanh}\left(\frac{\beta V}{2}\right),
\eeq

and 
\beq
K\left(x\right)=\int_{0}^{\pi/2} \frac{1}{\sqrt{1-x \sin^2 (\theta)}} d\theta,  
\eeq
is the complete elliptic integral of the second kind. From the above internal energy, we can infer the heat capacity by 
\beq
c_{\mu}=\frac{1}{N}\frac{dU}{dT}.
\eeq
We have compared the results of our simulations for $V'=0$ and for a system size of $30\times30$ with the above expressions in Fig.~\ref{fig:exactsol}. Unsurprisingly, both quantities show excellent agreement with the exact result down to $T\sim0.02V$. Beyond these lowest temperatures, the highly correlated nature of the configurations leads to a significant slowing down of the updates, thereby leading to a lack of appreciable energy fluctuations for our MC calculation. However, these issues can be mitigated, in principle, by performing replica exchange MC simulation \cite{ReplicaExc}. In this method, the different MC simulations are performed in parallel at slightly different temperatures and their configurations can be exchanged in accordance with the Metropolis acceptance rule, thereby reducing the correlation time. However, this lies beyond the scope of our present needs. 

\begin{figure}
\centering
\includegraphics[width=1\linewidth]{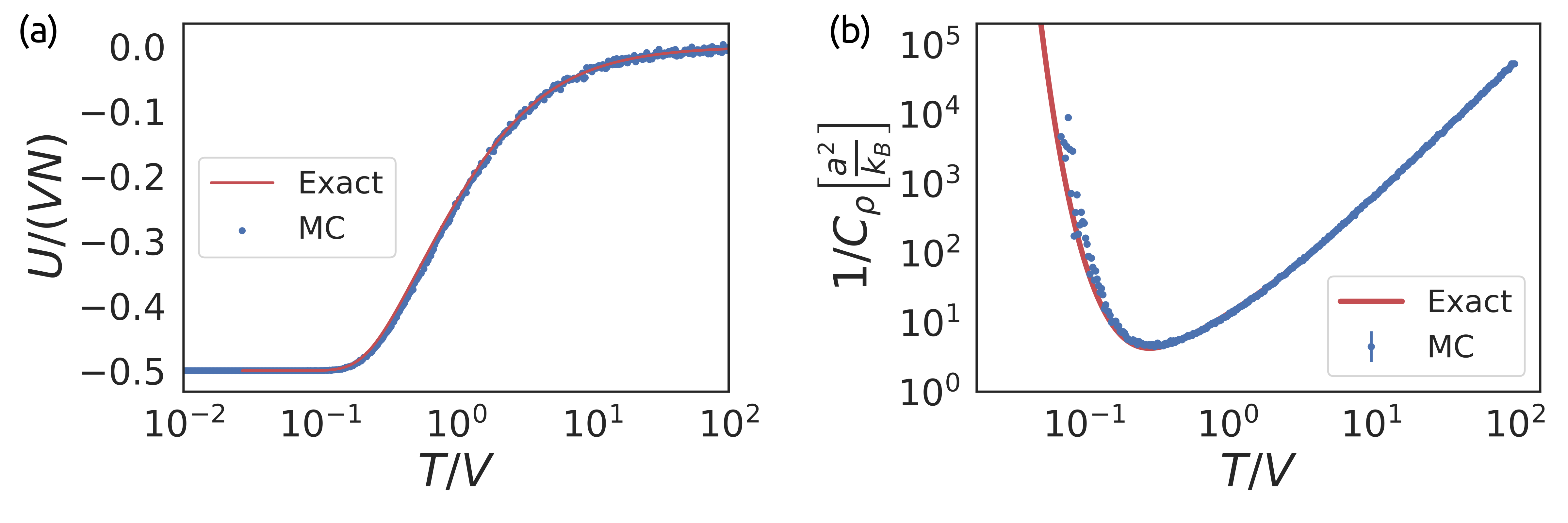}
\caption{\textsf{Comparison between the exact solution by Wannier and MC calculation for a system size of $30\times30$.}}
\label{fig:exactsol} 
\end{figure}

We can characterize the energetics associated with the exact solution at zero field by starting from the $\sqrt{3}\times\sqrt{3}$ CDW. First, consider the internal energy of this state, which
can be calculated by noticing that there are a total of $6$ bonds per
site. Since each bond is shared between two sites, there are
only $6N/2=3N$ different bonds in total. However, in the $\sqrt{3}\times\sqrt{3}$
CDW, $3$ of them are paired per site for a third of the sites. From
this, we conclude that there are a total of $N/2$ unfrustrated bonds,
each contributing an energy $-\left|J\right|$. As such, the total
energy of this state is $U=-N\left|J\right|/2$. This value precisely
coincides with the $T\rightarrow0$ limit of the internal energy of
the exact solution above. As such, this $\sqrt{3}\times\sqrt{3}$
CDW configuration is included in the ground state manifold. 

However, the above $\sqrt{3}\times\sqrt{3}$ CDW is not the only possible
ground state configuration. It can be easily verified that flipping the occupation
of any site in the unoccupied sublattices leaves the energy invariant.
In fact, the same pattern can occur in any of three different sublattices
and with either occupied or empty sites at the centers of the `star',
leaving the energy unchanged. Moreover, for $\mu=0$, the energetic
cost of phase boundaries between these regions can be made to vanish.
Consequently, there is a proliferation of phase boundaries which do
not carry an energetic cost and that prevent ordering.

The nature of the massively degenerate ground state manifold has implications for the qualitative features of the spin susceptibility. At $T=0$, the addition of any small external field ($\mu$) will add an energetic cost for a site to be either occupied or empty (depending on the sign of $\mu$). This, in turn, imposes
an energy penalty for the phase boundaries, stabilizing an
ordered state. The drastic change in occupation associated with the inclusion
of an infinitesimal $\mu$ results in a diverging spin susceptibility at
$T=0$.

\end{appendix}

\bibliography{nfl}

\end{document}